\documentclass[aps,showpacs,preprintnumbers,amsmath,amssymb,superscriptaddress,floatfix,nofootinbib]{revtex4}
\usepackage{lipsum}
\usepackage{slashed}
\usepackage{graphicx}
\usepackage{epsfig}
\usepackage{epstopdf}
\usepackage{hyperref}
\usepackage{amsmath}
\usepackage{amsfonts}
\usepackage{amssymb}
\usepackage{setspace}
\usepackage{amsfonts,color}
\usepackage{caption}
\usepackage{natbib}
\usepackage{subfig}
\usepackage{color}
\usepackage{bbold}
\usepackage{bm}
\usepackage{tabularx}
\usepackage{float}
\usepackage{tikz}
\usepackage[compat=1.1.0]{tikz-feynman}

\def\w2{\tilde w^2}
\def\ws2{1}

\begin{document}

\title{Charmed baryon productions in proton-antiproton collisions \\ in effective Lagrangian and Regge approaches}

\author{Thanat Sangkhakrit}
\email[]{tanattosan@outlook.jp}
\affiliation{School of Physics and Center of Excellence in High Energy Physics \& Astrophysics,
Suranaree University of Technology, Nakhon Ratchasima 30000, Thailand}
\affiliation{Research Center for Nuclear Physics,
Osaka University, Ibaraki, 567-0048, Japan}

\author{Sang-In Shim}
\affiliation{Research Center for Nuclear Physics,
Osaka University, Ibaraki, 567-0048, Japan}

\author{Yupeng Yan}
\email[]{yupeng@g.sut.ac.th}
\affiliation{School of Physics and Center of Excellence in High Energy Physics \& Astrophysics,
Suranaree University of Technology, Nakhon Ratchasima 30000, Thailand}

\author{Atsushi Hosaka}
\email[]{hosaka@rcnp.osaka-u.ac.jp}
\affiliation{Research Center for Nuclear Physics,
Osaka University, Ibaraki, 567-0048, Japan}
\affiliation{Advanced Science Research Center,
Japan Atomic Energy Agency, Tokai, Ibaraki, 319-1195 Japan}

\date{\today}
\begin{abstract}
\indent \indent 
Strange and charmed baryon productions from proton-antiproton collisions are studied in the effective Lagrangian and Regge approaches. 
We include only the $t$-channel dynamics which is dominant for the diffractive region that is relevant in the present discussions. 
For strangeness productions, the coupling constants for $K$ meson are determined by $SU(3)$ relations while other unknown parameters for $K^{*}$ couplings and for form factors are fixed by existing observed data for strangeness productions. 
By extrapolating the amplitudes for strangeness productions to those for charm, we predict charm production cross sections. 
It turns out that the total cross sections are $10^{4}$ to $10^{5}$ smaller than those of strangeness productions, 
depending on the final states. 
By using the known total cross sections for strangeness, $\sim 10^{2} \mu b$ for $\Lambda \bar{\Lambda}$ and, $\sim 10^{1} \mu b$ for $\Sigma \bar{\Lambda}$ the total cross sections for charm productions are predicted to be $\sim 10^{-2} \mu b$ for $\Lambda_{c} \bar{\Lambda}_c$, $\sim 10^{-3} \mu b$ for $\Sigma_{c} \bar{\Lambda}_c$, and $\sim 10^{-4} \mu b$ for $\Sigma_{c} \bar{\Sigma}_c$. Our results can be tested in the future experiments at $\bar{\text{P}}$ANDA.
\end{abstract}


\maketitle
\section{Introduction}
\indent \indent 
Physics of hadrons containing charm quarks, or in general heavy quarks, has been one of actively studied subjects in hadron physics 
since the first observations of $J/\psi$ meson in 1974 \cite{Augustin:1974xw,Aubert:1974js} 
and of the charmed baryon states $(\Sigma_{c}, \Lambda_{c})$ in 1975 \cite{Cazzoli:1975et}. 
In particular, in the 21st century various hadrons that are often called exotic hadrons 
have been observed by Belle, BABAR, BESIII, and LHCb collaborations \cite{Choi:2003ue,Aubert:2004ns,Aubert:2005rm,Abe:2007jna,Choi:2007wga,Hosaka:2016pey,Ablikim:2013mio,Liu:2013dau,Ablikim:2013wzq,Aaij:2013zoa,Aaij:2014jqa}. 
Accordingly, theoretical studies have been performed extensively in QCD inspired approaches such as the quark model \cite{Micu:1968mk,Godfrey:1985xj,Maiani:2004vq,Ebert:2005nc,Limphirat:2010zz,Xu:2020ppr}, 
models of chiral and heavy quark symmetries \cite{Gupta:1994mw,Ebert:1997nk,Glozman:2003bt,Nowak:2003ra,AlFiky:2005jd,Liu:2006jx}, non-relativistic QCD \cite{Brambilla:1999xf,Braguta:2005kr},  and also by the first principle calculations of lattice QCD \cite{Isgur:1984bm,Chen:2000ej,Okamoto:2001jb,Liao:2002rj,McNeile:2002az,Chiu:2005ey,Chiu:2006hd} (See Ref.\cite{Swanson:2006st,Brambilla:2010cs} for reviews).
These studies are mostly focused on the structure and decays, and much knowledge of them have been accumulated.
In contrast, their productions are less investigated, as they are in many cases considered to be complicated
due to their rather inclusive nature.
Physically, the production mechanism of heavy quarks from the light quarks in the initial state is 
an interesting and the least understood issue.
Quantitative descriptions of such heavy quark productions especially in exclusive processes 
provide better understanding not only of non-perturbative dynamics of QCD
but also of the structure of the various hadrons.  

As an example of such an exclusive process a pion induced reaction for charmed baryon production was studied many years ago at Brookhaven \cite{Christenson:1985ms}, which however reported only null results.  
An updated experiment is planned at J-PARC and the construction of the facility is on-going \cite{e50}, and the corresponding 
theoretical studies have been also carried out \cite{Kim:2015ita,Kim:2015omh,Shim:2019uus,Shim:2019yxn}.
In this paper we study another process induced by anti-protons.  
This is a planned experiment at GSI as $\bar{\text{P}}$ANDA project of FAIR, 
providing another promising reactions to produce charmed baryons.
At the same time, they can also produce strange hyperons, and so 
we can study systematically both strange and charm productions.
An exclusive set up for identifying the processes such as 
\begin{equation}
p  \bar p \to \Lambda_c   \bar \Lambda_c, \  \Sigma_c   \bar \Lambda_c, \  \Lambda_c   \bar \Lambda_c^*, ...
\end{equation}
where stared ones are resonances provides an ideal opportunity for the study of various baryons including the ground and resonant states.  


\indent \indent 
By now several models for charm productions have been proposed and cross sections have been computed.  
In Ref.\cite{Kroll:1988cd}, various charm production cross sections have been estimated in the quark-diquark picture.
In this model, charmed hadrons are produced via the interaction between the active constituents (quark or diquark) of the initial states. 
This approach has some similarity to the handbag approach used in Ref.\cite{Goritschnig:2009sq}. 
In this case, the transition amplitude of the reaction $p\bar{p} \to \Lambda_{c}\bar{\Lambda}_{c}$ is computed in terms of the amplitude for the hard subprocess (i.e., $u\bar{u} \to c\bar{c}$) 
and the soft hadronic matrix elements for $p \to \Lambda_{c}$ and $\bar{p} \to \bar{\Lambda}_{c}$ transitions. 
Quark-Gluon String Model (QGSM) and Regge approach are employed by several authors in Ref.\cite{Kaidalov:1994mda, Titov:2008yf, Khodjamirian:2011sp}. 
In this model, the annihilation of $q\bar{q}$ pair from the initial states is followed by the formation of the intermediate string, 
then the observed charmed hadrons are consequently produced from the string fission. 
Meson-exchange framework was employed in Ref.\cite{Haidenbauer:1991kt,Haidenbauer:2016pva} to compute $\Lambda_{c}\bar{\Lambda}_{c}$ production cross sections. 
This model was developed from J$\ddot{\text{u}}$lich meson-baryon model, 
which is originally employed to compute the cross sections for the reaction $p\bar{p} \to \Lambda \bar{\Lambda}$ within the coupled-channel framework.
Effective Lagrangian approach with coupling constants from $SU(4)$ symmetry is employed in Ref.\cite{Shyam:2014dia} to compute various charm productions, which are produced via $D$ and $D^{*}$ meson-exchange processes.
From the various models we mentioned earlier,
a strong model dependence is observed as their predicted charm production cross sections are different by several orders. \newline
\indent \indent In our present work, we study strangeness and charm productions in effective Lagrangian and Regge approaches. 
We start with differential cross sections for strangeness productions, and then we extrapolate the results to charm ones. 
In the first two sections II and III, these production cross sections will be computed in the framework of effective Lagrangian approach, and then in the Regge approach. 
Then, total cross sections from each method will be presented and then compared with the experimental data in section IV. We summarize our paper in section V.
\label{sec-1}

\section{Effective Lagrangian Approach}
\label{sec-2}
\indent \indent In this section, effective Lagrangian method is employed to study differential cross sections of strangeness productions from proton-antiproton scattering. 
This method allows us to describe hadron-hadron scattering near the production threshold. 
First we show the relevant effective Lagrangians that we use and then Feynman amplitudes for the production processes at the tree level are given. 
Differential cross sections for various strange baryons are presented. 
By replacing strange mesons and hyperons with the charm ones, differential cross sections for charmed baryons are predicted.
\subsection{Effective Lagrangians and Feynman amplitudes}
\begin{figure}[H]
    \centering
    \includegraphics[scale=1]{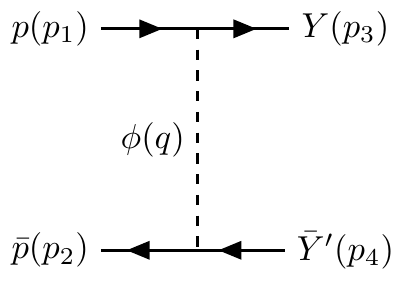}
    \caption{Tree-level diagram for the reaction $p\bar{p} \to Y\bar{Y}^{\prime}$}
    \label{treelev}
\end{figure}

\indent \indent To start with, let us first discuss strangeness productions, 
and then later we shall discuss charm productions. 
Here, we assume that cross sections for such processes are dominated by t-channel exchanges, this assumption is reasonable for the scattering at energy that are sufficient above the threshold. 
The relevant Feynman diagram is displayed in Fig.\ref{treelev}, 
where $Y$ and $\bar{Y}^{\prime}$ denote a produced hyperon ($\Lambda^{0}$ or $\Sigma^{0}$) and its antiparticle ($\bar{\Lambda}^{0}$ or $\bar{\Sigma}^{0}$). 
An exchanged strange meson ($K^{+}$ or $K^{*+}$) is denoted by $\phi$. 
The momenta of the incoming proton and antiproton are denoted by $p_{1}$ and $p_{2}$ while $q$, $p_{3}$, and $p_{4}$ are those of the exchanged strange meson, outgoing hyperon, and antihyperon, respectively.   
\indent \indent The relevant effective Lagrangians for $KNY$ and $K^{*}NY$ vertices are given as,
\begin{align}
\mathcal{L}_{KNY} &= \dfrac{g_{KNY}}{m_{N}+m_{Y}} \bar{N}\gamma^{\mu}\gamma_{5}Y \partial_{\mu}K + \text{H.c.,} \label{Lp}\\
\mathcal{L}_{K^{*}NY} &= -g_{K^{*}NY}\bar{N}\left[\gamma^{\mu}Y - \dfrac{\kappa_{K^{*}NY}}{m_{N}+m_{Y}}\sigma^{\mu\nu}Y\partial_{\nu}\right]K_{\mu} + \text{H.c.,}
\label{Lv}
\end{align}
where $Y$ stands for isoscalar state $\Lambda$ or isotriplet state $\bm{\Sigma} \cdot \bm{\tau}$. 
Here, the $\tau$ matrices operate to the isospin states of the nucleon and $K$ or $K^{*}$. 
The following axial-vector coupling constants with $K$ meson,
\begin{equation}
g_{KN\Lambda} = -13.5, \hspace{1 cm} g_{KN\Sigma} = 2.5,
\end{equation}
are derived from $SU(3)$ symmetry relations, where the coupling constants of baryon-baryon-meson vertices are represented in terms of the $\pi NN$ coupling constant and the parameter $\alpha$ \cite{Rijken:1998yy},
\begin{align}
\dfrac{g_{KN\Lambda}}{g_{\pi NN}} &= -\dfrac{1}{\sqrt{3}}\left(3-2\alpha\right), \\
\dfrac{g_{KN\Sigma}}{g_{\pi NN}} &= -(1-2\alpha).
\end{align}
Here $g_{\pi NN} = 13$ and $\alpha = \dfrac{D}{F+D} \sim \dfrac{3}{5}$ determined from the analysis of baryon-baryon interactions in a one-boson-exchange-potential approach \cite{Nagels:1978sc} is employed.
This numerical value agrees with the one obtained from the $SU(6)$ prediction \cite{Pais:1966} and large-$N_{c}$ analysis \cite{Dashen:1994qi}.  
To express the finite size of hadrons, following form factors are utilized in our calculations,
\begin{align}
F\left(t\right) &= k^{2}\dfrac{\Lambda^{4}}{\Lambda^{4} + \left(t-m_{\phi}^{2}\right)^{2}}, \label{F}\\
F_{n}\left(t\right) &= k \left( \dfrac{\Lambda^{2}}{\Lambda^{2}-t} \right)^{n}, \hspace{0.5cm} \left(n=1,2\right), \label{dF}
\end{align}
where $t=q^{2}$ and $m_{\phi}$ is the mass of the exchanged meson. 
Here, the form factor in Eq.(\ref{F}) is multiplied to the whole Feynman amplitude (See  Refs.\cite{Kim:2015ita,Nam:2005uq,Haberzettl:1998aqi}), while the second one in Eq.(\ref{dF}) is at each vertex. 
Therefore, in the Feynman amplitude, the normalization constant appears in $k^{2}$ for both cases. 
Because there is always ambiguities in the use of form factors for such reaction processes, we will compare the results by using the three different form factors. \newline
\indent \indent There are four parameters in our model: vector coupling constant $g_{K^{*}NY}$, tensor coupling constant $\kappa_{K^{*}NY}$, normalization constant $k$, and cutoff parameter $\Lambda$. 
These parameters will be fixed by the existing observed data of relevant strangeness production cross sections. \newline
\indent \indent By applying Feynman rules to the effective Lagrangians in Eq.(\ref{Lp}) and Eq.(\ref{Lv}), Feynman amplitudes of strangeness productions with $K$ and $K^{*}$ exchanges are given by
\begin{align}
\mathcal{M}_{K} &= \dfrac{g_{KNY}g_{KN\bar{Y}^{\prime}}}{\left(m_{N}+m_{Y}\right)\left(m_{N}+m_{\bar{Y}^{\prime}}\right)} \Gamma_{N}\left(p_{1},s_{1},p_{3},s_{3}\right)  P_{K}\left(t\right) \Gamma_{\bar{N}}\left(p_{2},s_{2},p_{4},s_{4}\right), \label{FAP} \\
\mathcal{M}_{K^{*}} &= g_{K^{*}NY}g_{K^{*}N\bar{Y}^{\prime}} \Gamma_{N, \mu}\left(p_{1},s_{1},p_{3},s_{3}\right) P_{K^{*}}^{\mu \nu} \left(t\right) \Gamma_{\bar{N},\nu}\left(p_{2},s_{2},p_{4},s_{4}\right),
\label{FAV}
\end{align}
where
\begin{align}
\Gamma_{N}\left(p_{1},s_{1},p_{3},s_{3}\right) &= \bar{u}_{Y}\left(p_{3},s_{3}\right)\slashed{q}\gamma_{5}u_{N}\left(p_{1},s_{1}\right), \\
\Gamma_{\bar{N}}\left(p_{2},s_{2},p_{4},s_{4}\right) &=  \bar{v}_{N}\left(p_{2},s_{2}\right)\slashed{q}\gamma_{5}v_{\bar{Y}^{\prime}}\left(p_{4},s_{4}\right),\\
\Gamma_{N, \mu}\left(p_{1},s_{1},p_{3},s_{3}\right) &= \bar{u}_{Y}\left(p_{3},s_{3}\right) \left[ \left(1+\kappa_{K^{*}NY}\right)\gamma_{\mu} - \kappa_{K^{*}NY}\dfrac{\left(p_{1}+p_{3}\right)_{\mu}}{m_{N}+m_{Y}} \right] u_{N}\left(p_{1},s_{1}\right),\\
\Gamma_{\bar{N},\nu}\left(p_{2},s_{2},p_{4},s_{4}\right) &= \bar{v}_{N}\left(p_{2},s_{2}\right) \left[ \left(1+\kappa_{K^{*}N\bar{Y}^{\prime}}\right)\gamma_{\nu} + \kappa_{K^{*}N\bar{Y}^{\prime}}\dfrac{\left(p_{2}+p_{4}\right)_{\nu}}{m_{N}+m_{\bar{Y}^{\prime}}} \right] v_{\bar{Y}^{\prime}}\left(p_{4},s_{4}\right).
\end{align}
The Feynman propagators for $K$ and $K^{*}$ mesons are defined by
\begin{align}
P_{K}\left(t\right) &= \dfrac{i}{t-m_{K}^{2}},\\
P_{K^{*}}^{\mu \nu}\left(t\right) &= \dfrac{i \left(-g^{\mu \nu} + q^{\mu}q^{\nu}/m_{K^{*}}^{2}\right)}{t-m_{K^{*}}^{2}}.
\end{align}
The total Feynman amplitude is therefore written as
\begin{equation}
\mathcal{M}_{p\bar{p} \to Y\bar{Y}^{\prime}} = \begin{cases}
   \mathcal{M}_{K}F_{K} + \mathcal{M}_{K^{*}}F_{K^{*}}, \\
    \mathcal{M}_{K}F_{n, K}^{2} + \mathcal{M}_{K^{*}}F_{n, K^{*}}^{2}.
  \end{cases} 
\end{equation}
Here $F_{(K,K^{*})}$ and $F_{n, (K,K^{*})}$ are form factors in Eq.(\ref{F}) and Eq.(\ref{dF}) with $K$ or $K^{*}$, respectively. 
Differential cross section as a function of a momentum transfer $t$ is computed from,
\begin{align}
\dfrac{d\sigma}{dt} &= \dfrac{1}{64\pi\left(p_{cm}\right)^{2}s}\left\langle \left|\mathcal{M}\right|^{2} \right\rangle,
\label{sigmadt}
\end{align}
where $\left\langle \left|\mathcal{M}\right|^{2} \right\rangle = \dfrac{1}{4}\sum_{s_{3},s_{4}} \left|\mathcal{M}\right|^{2}$ and $p_{cm}$ is the relative momentum of the $p$ and $\bar{p}$ in the initial state in the center of mass frame. \\
\begin{figure}[H]
    \centering
    \includegraphics[scale=1]{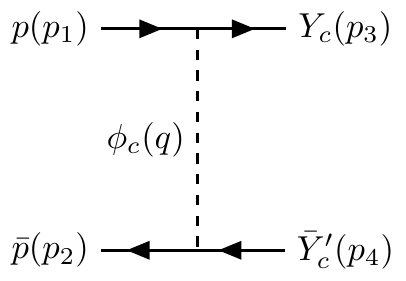}
    \caption{Tree-level diagram for the reaction $p\bar{p} \to Y_{c}\bar{Y}^{\prime}_{c}$}
    \label{treelevc}
\end{figure}
\indent \indent  Now we proceed to the discussion of charm productions. The relevant Feynman diagram is displayed in Fig.\ref{treelevc} where $Y_{c}$ and $\bar{Y}_{c}^{\prime}$ denote a produced charmed baryon ($\Lambda_{c}^{+}$ or $\Sigma_{c}^{+}$) and its antiparticle ($\bar{\Lambda}_{c}^{-}$ or $\bar{\Sigma}_{c}^{-}$). Whereas an exchanged charmed meson ($\bar{D}^{0}$ or $\bar{D}^{*0}$) is denoted by $\phi_c$. Here, the corresponding amplitudes are obtained by replacing strange mesons and hyperons in Eq.(\ref{FAP}) and Eq.(\ref{FAV}) with the charm ones. In principle, the coupling constants for charm hadrons and for strange ones should be different. However, if strange and charm quarks are sufficiently heavy. the same set of coupling constants can be applied to the vertices in Fig.\ref{treelevc}. 

\subsection{Results for Strangeness Productions}
\begin{figure}[H]
\centering
\includegraphics[width=5.5cm]{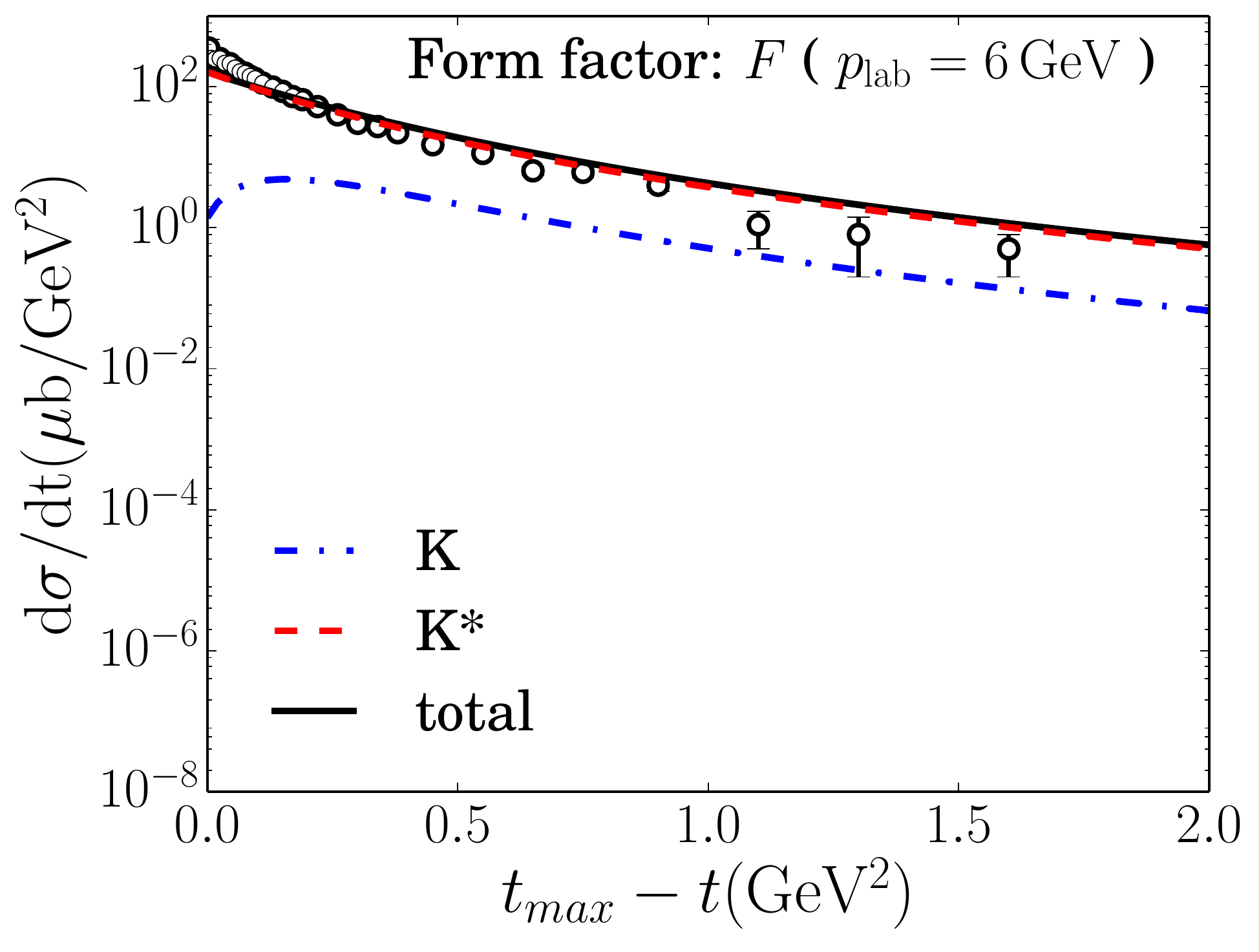}\includegraphics[width=5.5cm]{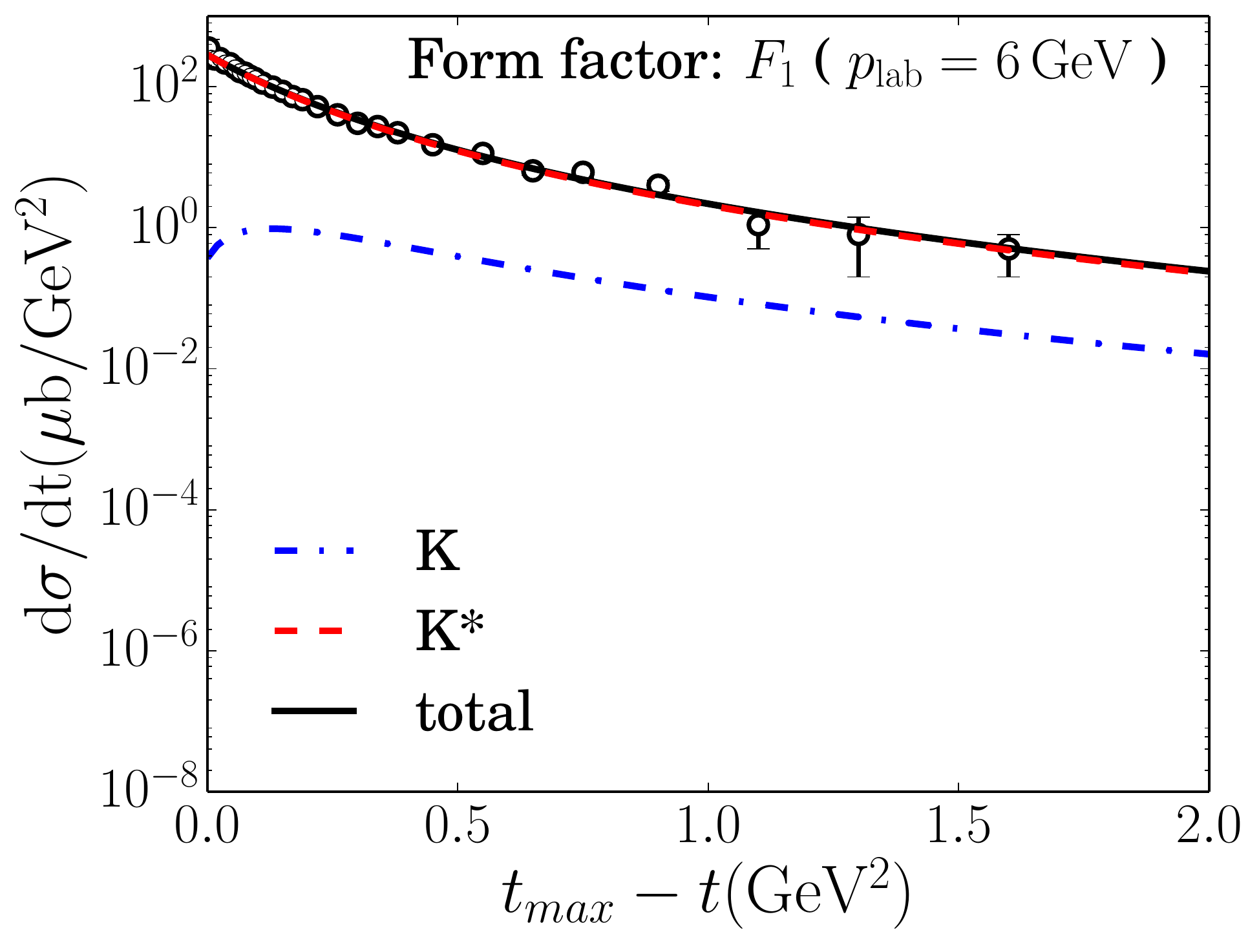}
\includegraphics[width=5.5cm]{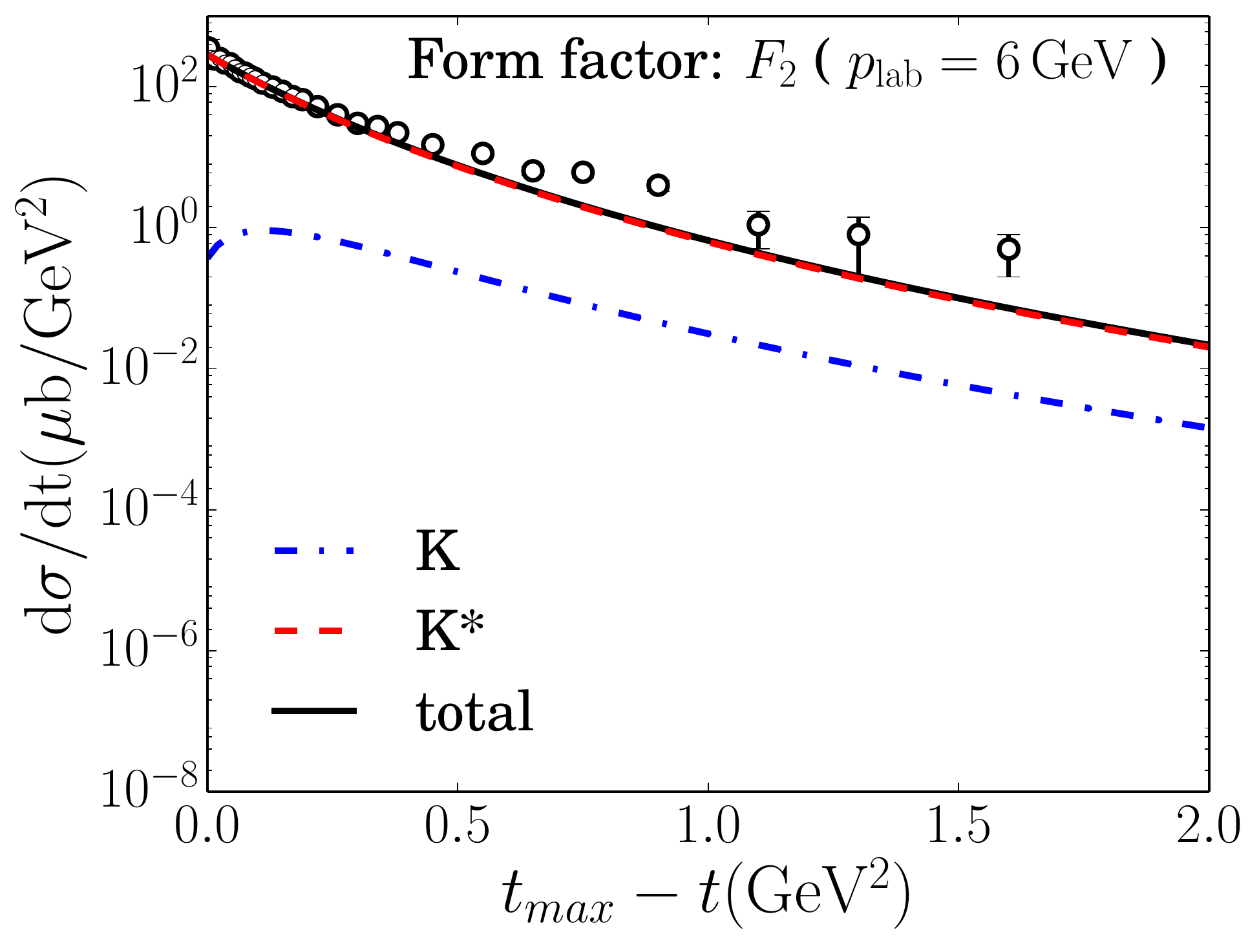}
\caption{Differential cross sections for $p\bar{p} \to \Lambda\bar{\Lambda}$ reaction at $p_{lab} = 6 \text{ GeV}$. The circles denote the experimental data taken from Ref.\cite{Becker:1978kk}.}
\label{LLEFT}
\end{figure}

\begin{figure}[H]
\centering
\includegraphics[width=5.5cm]{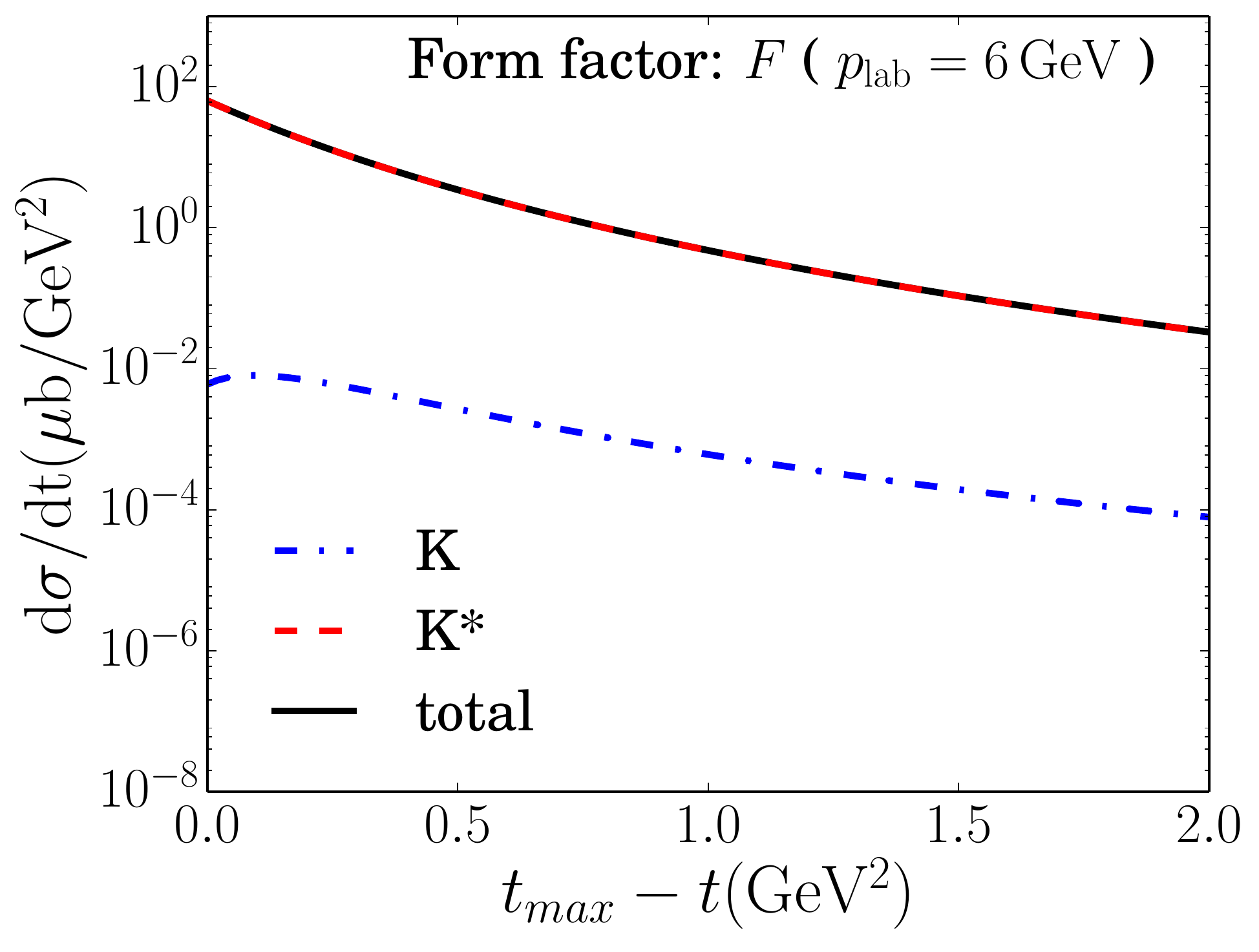}\includegraphics[width=5.5cm]{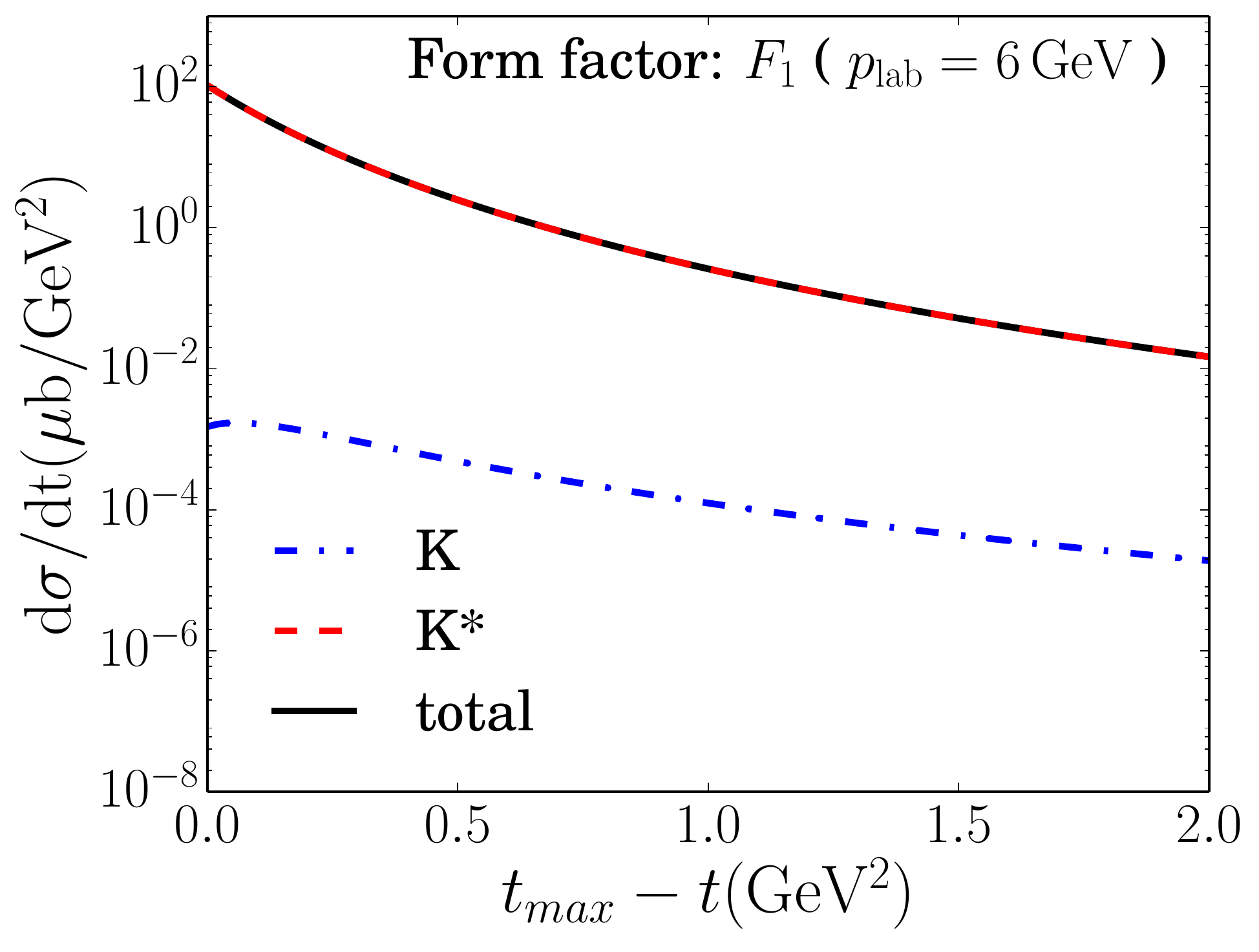}
\includegraphics[width=5.5cm]{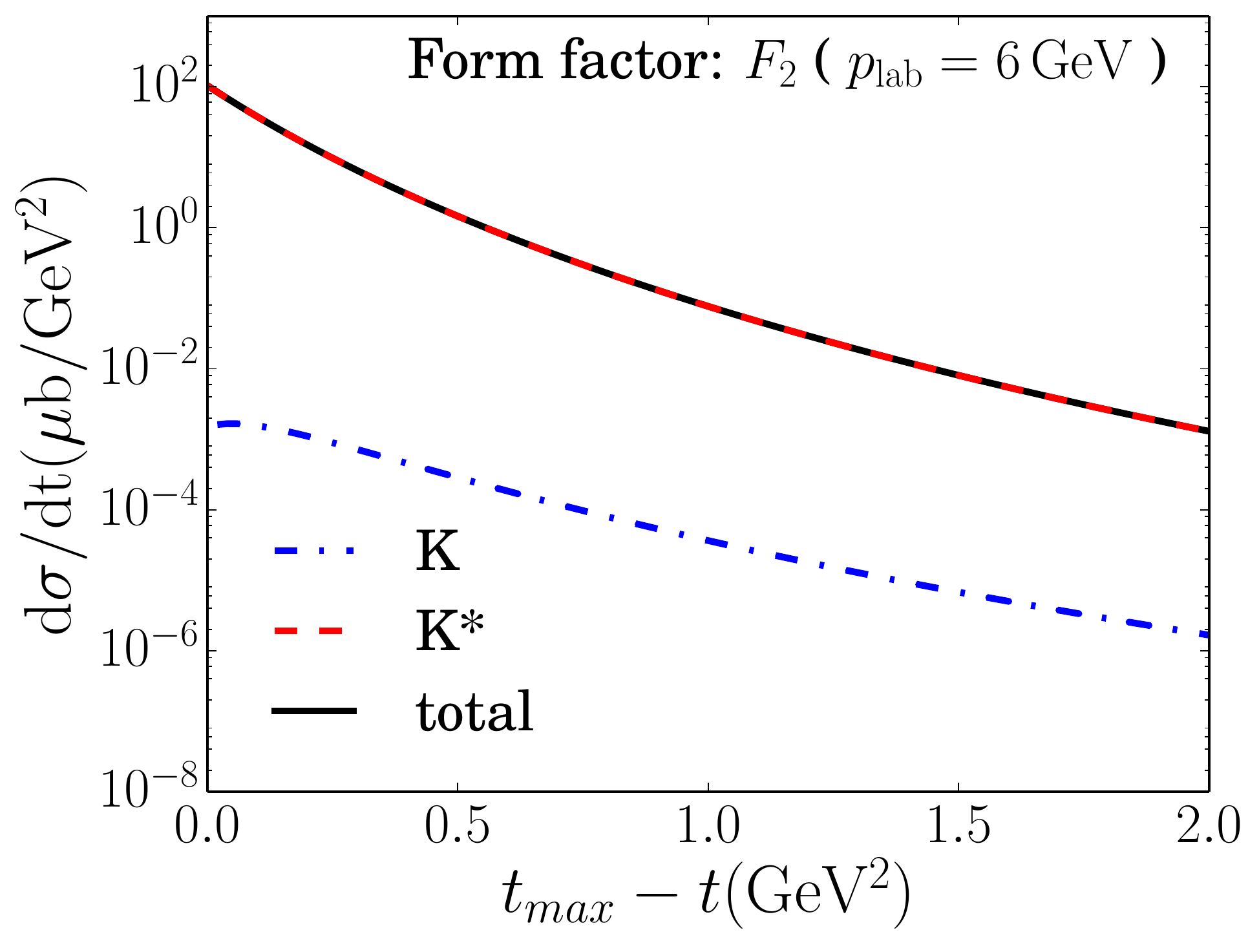}
\caption{Differential cross sections for $p\bar{p} \to \Sigma\bar{\Sigma}$ reaction at $p_{lab} = 6 \text{ GeV}$.}
\label{SSEFT}
\end{figure}

\indent \indent In the following subsection, differential cross sections for strange baryons as functions of $t_{max}-t$ are presented. 
In the observed differential cross sections of the reactions $p\bar{p} \to \Lambda\bar{\Lambda}$ and $p\bar{p} \to \Sigma\bar{\Lambda}$ \cite{Becker:1978kk}, which exist only for $p_{lab} = 6 \text{  GeV}$, It seems that there are two components: the steeper component in the forward angle region and the less steeper one at a larger angle region. 
In our present case, the first component of the experimental data is much important due to the $t$-channel dominance. 
At a given energy, the value of $t$ varies from $t_{min}$ and $t_{max}$(i.e., $t_{max}-t$ varies from $0$ to $t_{max}-t_{min}$). 
The momentum transfer at the forward angle productions $t_{max}$ and at the backward angle productions $t_{min}$ of the reaction $p\bar{p} \to Y\bar{Y}^{\prime}$ are written as,
\begin{equation}
t_{max}^{min} = m_{N}^{2} + m_{Y}^{2} - \dfrac{1}{2s}\left[ s \left(s+m^{2}_{Y}-m^{2}_{\bar{Y}^{\prime}}\right) \pm \sqrt{s \left(s-4m_{N}^{2}\right)\left(s-\left(m_{Y}+m_{\bar{Y}^{\prime}}\right)^{2}\right)\left(s-\left(m_{Y}- m_{\bar{Y}^{\prime}}\right)^{2}\right)} \right].
\end{equation}
We try to fix model parameters by comparing differential cross sections based on $(K + K^{*})$-meson exchange to the experimental data of $p \bar{p} \to \Lambda\bar{\Lambda}$ and $p \bar{p} \to \Sigma\bar{\Lambda}$ reactions. 
Then, differential cross sections for the reaction $p \bar{p} \to \Sigma\bar{\Sigma}$ will be predicted. 
For each reaction, differential cross sections based on $K$-meson exchange, $K^{*}$-meson exchange, and $(K + K^{*})$-meson exchange are displayed separately as well as the contributions from three different form factors. \newline
\indent \indent For the reaction $p \bar{p} \to \Lambda\bar{\Lambda}$, the following parameters,
\begin{align}
&g_{K^{*}N\Lambda} = -5.112, \hspace{0.25 cm} \kappa_{K^{*}N\Lambda} = 2.037, \nonumber \\
&F\left(t\right): \hspace{0.4cm} k = 0.46, \hspace{0.2cm} \Lambda = 0.63 \text{  GeV}, \nonumber \\
&F_{1}\left(t\right): \hspace{0.3cm} k = 0.285, \hspace{0.2cm} \Lambda = 0.7 \text{  GeV},\nonumber \\
&F_{2}\left(t\right): \hspace{0.3cm} k = 0.285, \hspace{0.2cm} \Lambda = 0.99 \text{  GeV},
\label{opLL}
\end{align}
are fixed such that differential cross sections with $(K + K^{*})$-meson exchange agree with the data as displayed in Fig.\ref{LLEFT}. 
By observing the results, the slope of the differential cross section is getting steeper from the left to the right panels. 
This means, different $t_{max} - t$ dependences are given by three different form factors.
The one with the form factor $F_1$ is in a reasonable agreement with the data in the whole $t_{max}-t$ region, while the deviation from the data in the finite angle region is observed from the ones with the form factors $F$ and $F_2$. 
However, these three different results are at least in a good agreement with the data near the forward angle region, which is consistent with the $t$-channel dominance in the diffractive region.
Moreover, we observe that $K^{*}$-meson exchange dominates the differential cross sections over the $K$-meson exchange. 
The ones with $K$-meson exchange are suppressed from the other ones with $K^{*}$-meson exchange by the factor $10^{-2}$ to $10^{-3}$. 
This $K^{*}$ dominance is implied by the model parameters in Eq.(\ref{opLL}). 
\begin{figure}[H]
\centering
\includegraphics[width=5.5cm]{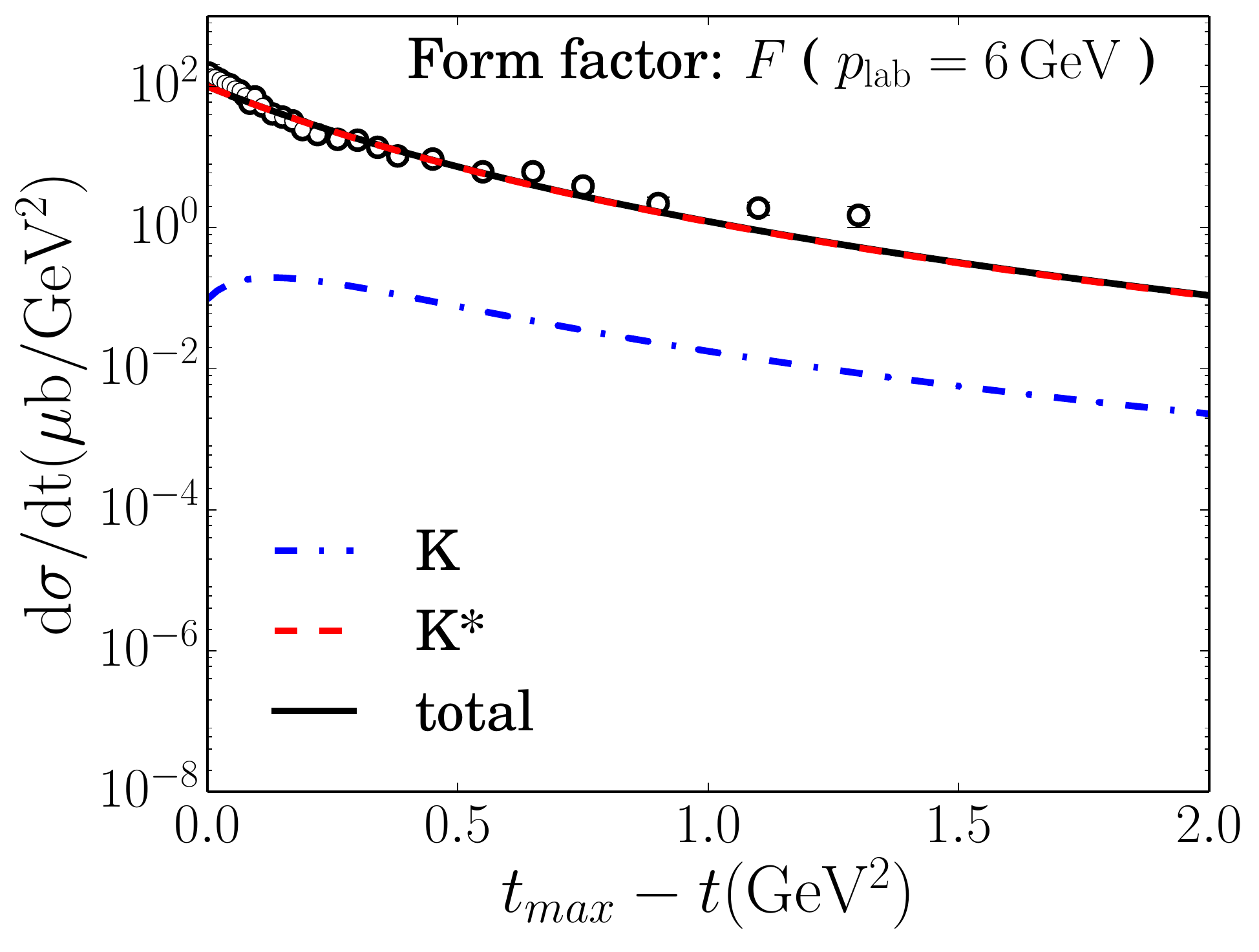}\includegraphics[width=5.5cm]{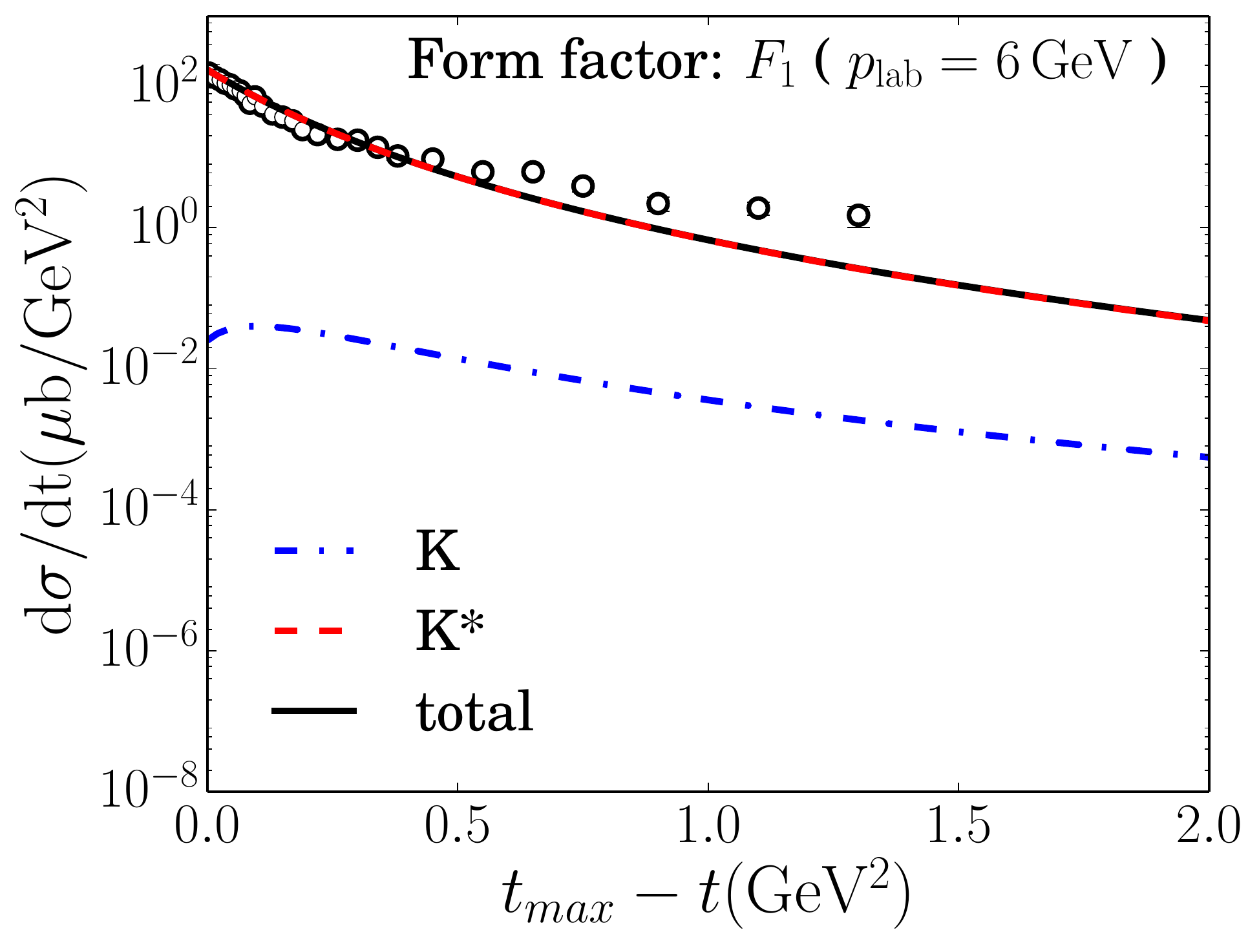}
\includegraphics[width=5.5cm]{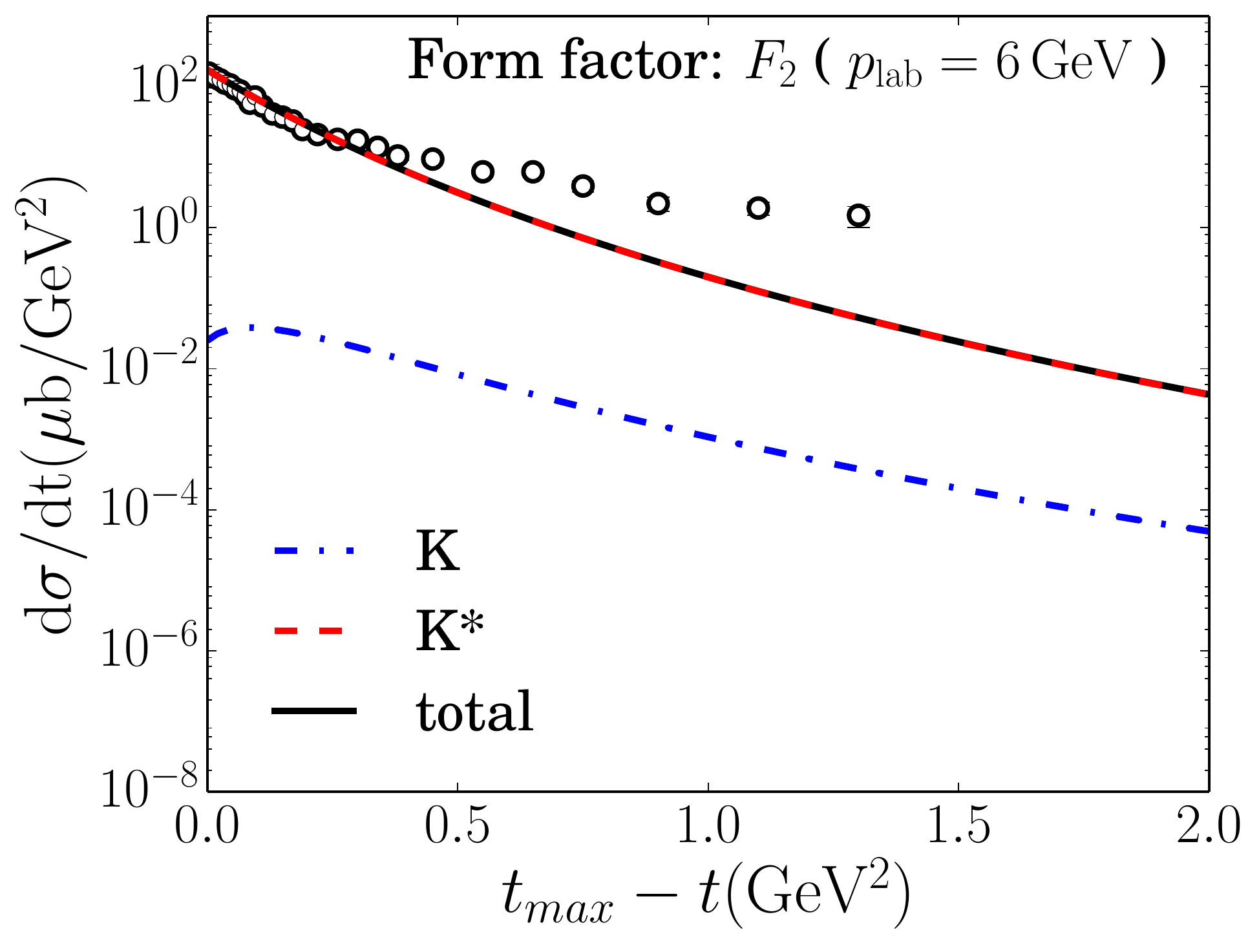}
\caption{Differential cross sections for $p\bar{p} \to \Sigma\bar{\Lambda}$ reaction at $p_{lab} = 6 \text{ GeV}$. The circles denote the experimental data taken from Ref.\cite{Becker:1978kk}.}
\label{SLEFT}
\end{figure}
\indent \indent Then, we repeat the same procedure to the reaction $p\bar{p} \to \Sigma\bar{\Lambda}$. In this case, $KN\Lambda$ and $K^{*}N\Lambda$ coupling constants determined from the reaction $p\bar{p} \to \Lambda\bar{\Lambda}$ are employed. The following coupling constants for the $K^{*}N\Sigma$-vertex,
\begin{equation}
g_{K^{*}N\Sigma} = -4.182, \hspace{1cm} \kappa_{K^{*}N\Sigma} = -0.6877,
\end{equation}
are fixed such that differential cross sections with $(K + K^{*})$-meson exchange agree with the data as displayed in Fig. \ref{SLEFT}. 
To reduce the number of model parameters, the same cutoff parameters are then applied to the reactions $p\bar{p} \to \Sigma\bar{\Lambda}$ and $p\bar{p} \to \Sigma\bar{\Sigma}$. 
The slope of the differential cross sections is getting steeper from the left to the right as what we have observed in $\Lambda\bar{\Lambda}$ production cross sections. 
In this case, the one with the form factor $F$ provides the best agreement with the data, while the deviation from the data is seen in the ones with the form factors $F_{1}$ and $F_{2}$. 
Anyway, our results agree with with the data near the forward angle region. 
Main contribution to the differential cross sections is again provided by $K^{*}$-meson exchange. 
The suppression factor between the cross sections with $K$ and $K^{*}$-meson exchanges is at the same order as the one obtained from $\Lambda\bar{\Lambda}$ production. 
\newline 
\indent \indent By employing model parameters determined from the two previous reactions, differential cross sections for the reaction $p\bar{p} \to \Sigma\bar{\Sigma}$ are predicted in Fig. \ref{SSEFT}. 
Here, the ones with $K$-meson exchange are suppressed from the other ones with $K^{*}$-meson exchange by the factor $10^{-5}$, which is obviously larger than what we have seen from $\Lambda\bar{\Lambda}$ and $\Sigma\bar{\Lambda}$ production cross sections. \newline
\indent \indent In summary, the agreement between differential cross sections and the data for strangeness productions is obtained if $K^{*}$-meson exchange dominates the differential cross sections at the given energy. 
This requirement is fulfilled by employing a suitable set of parameters such that the $t_{max} - t$ dependence and absolute values of the observed differential cross sections are reproduced.
The model parameters which we have fixed with the data will be employed to predict the differential cross sections for charm productions in the next subsection.
\subsection{Results for Charm Productions}
\begin{figure}[H]
\centering
\includegraphics[width=5.5cm]{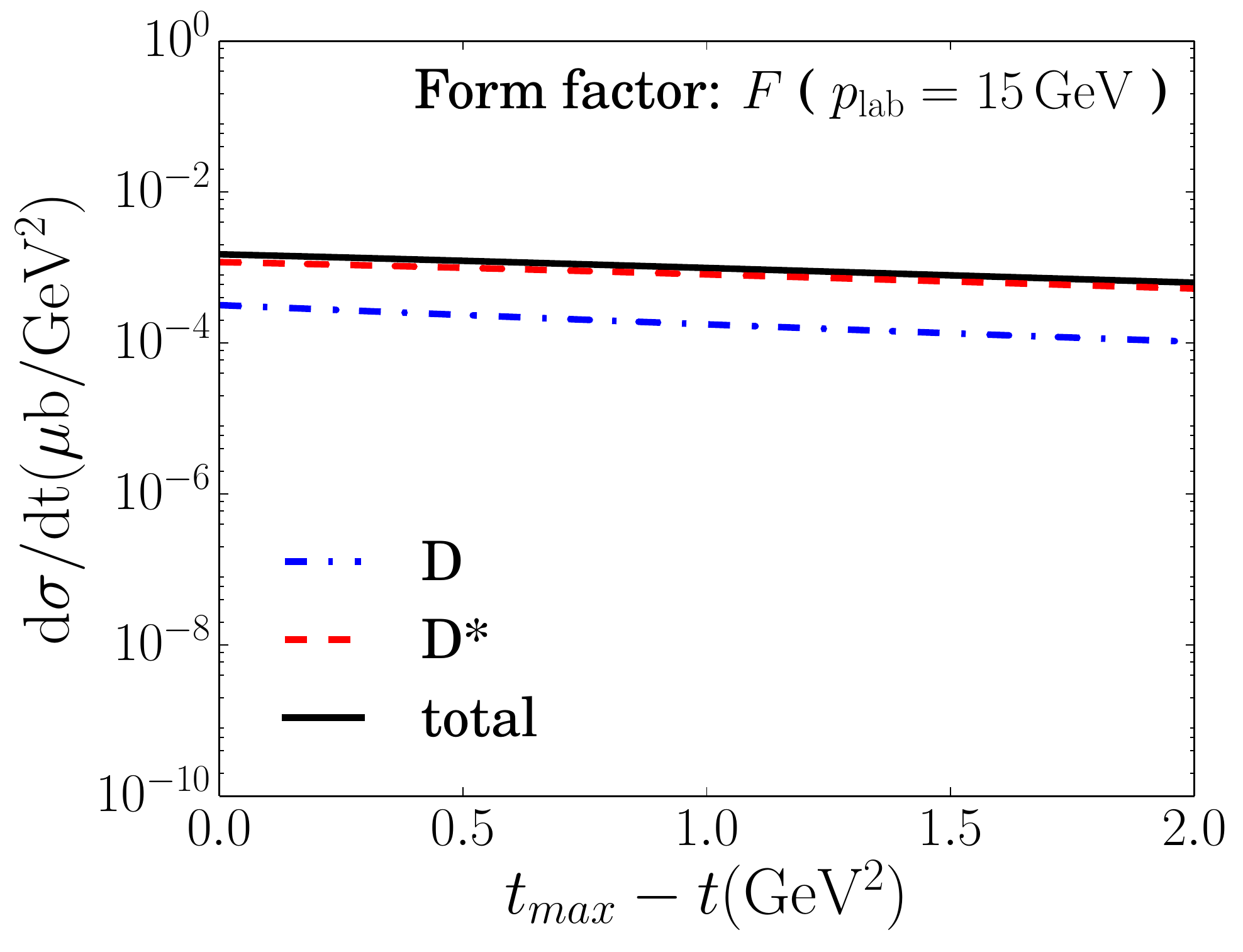}\includegraphics[width=5.5cm]{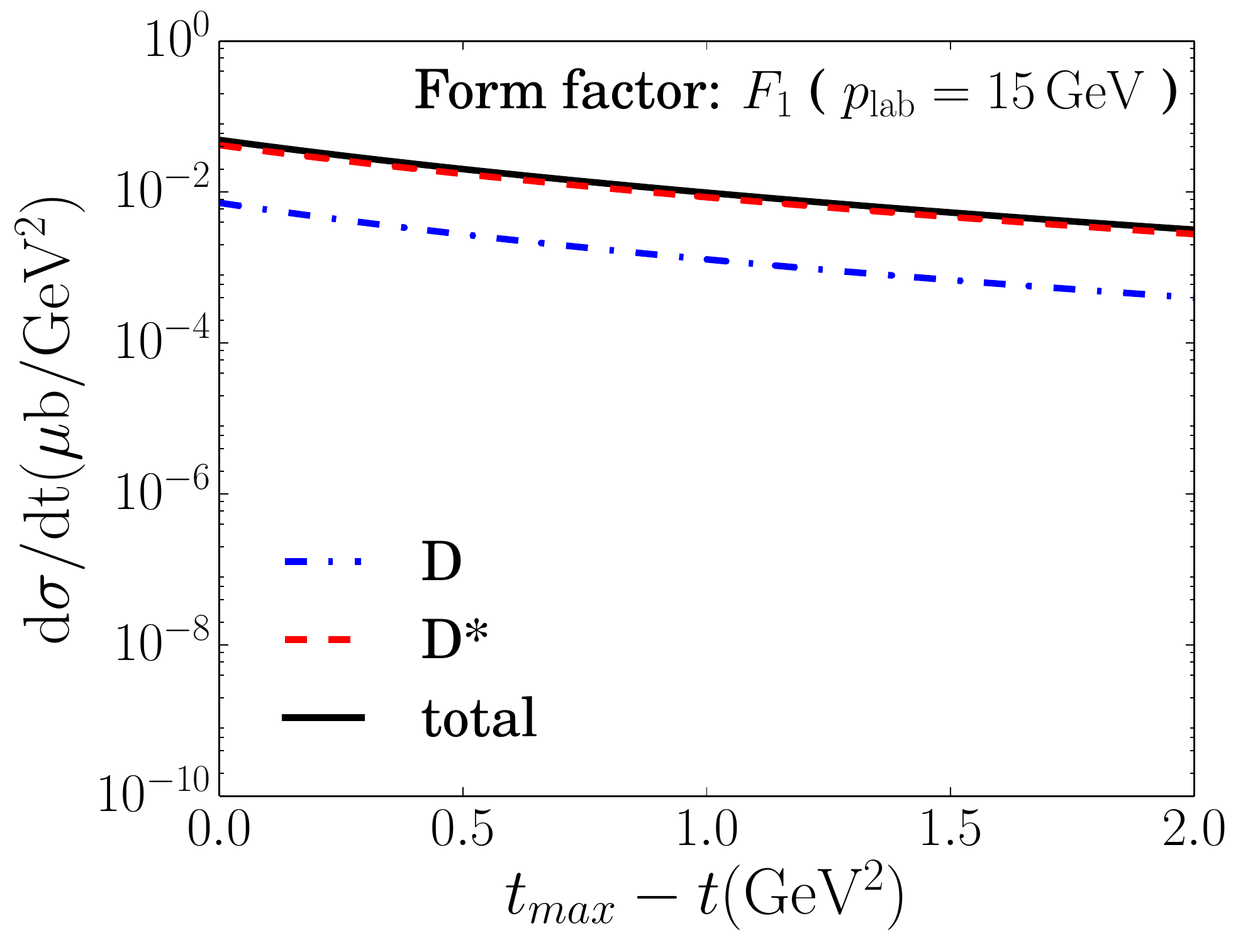}
\includegraphics[width=5.5cm]{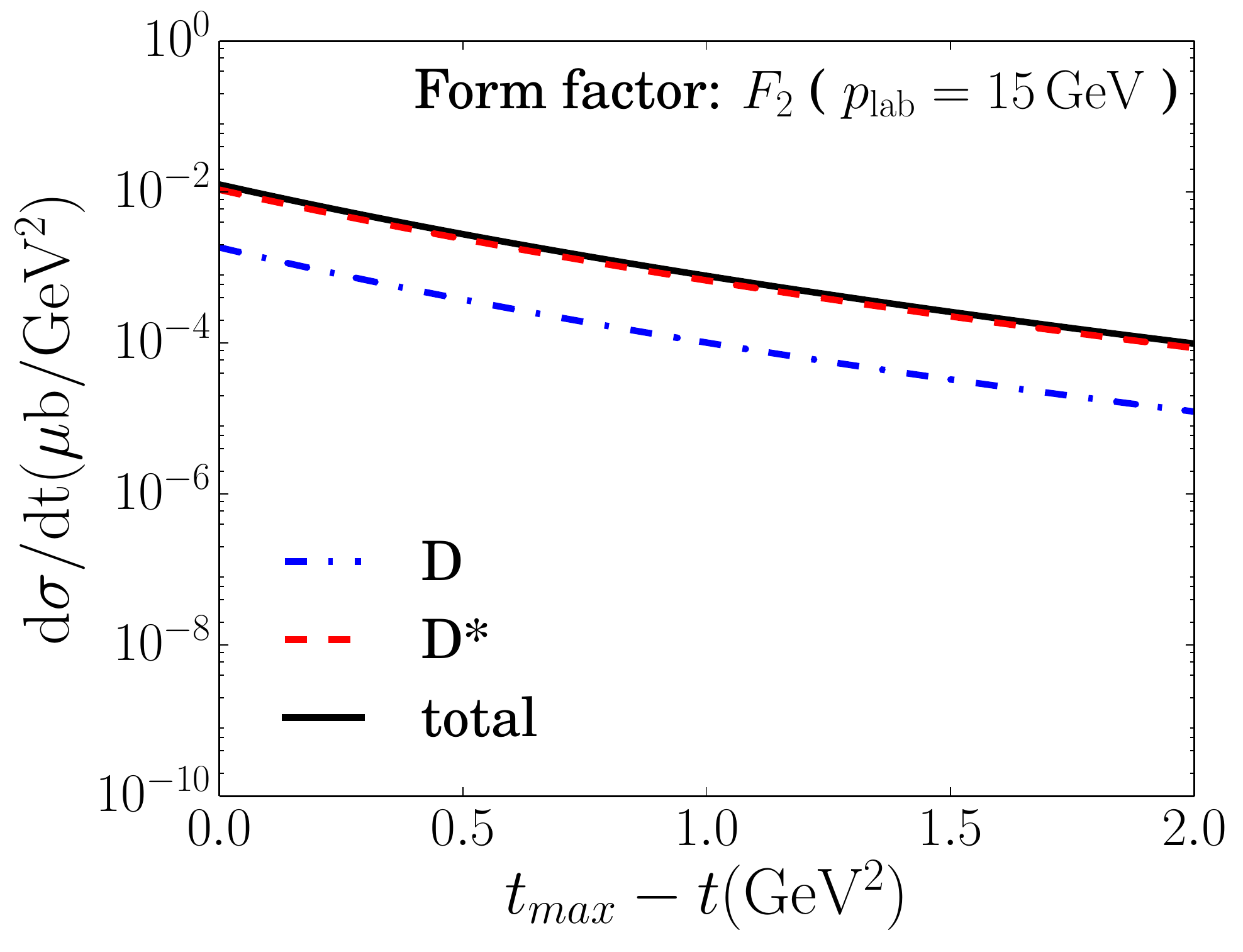}
\caption{Differential cross sections for $p\bar{p} \to \Lambda_{c}\bar{\Lambda}_{c}$ reaction at $p_{lab} = 15 \text{ GeV}$.}
\label{LLcEFT}
\end{figure}
\begin{figure}[H]
\centering
\includegraphics[width=5.5cm]{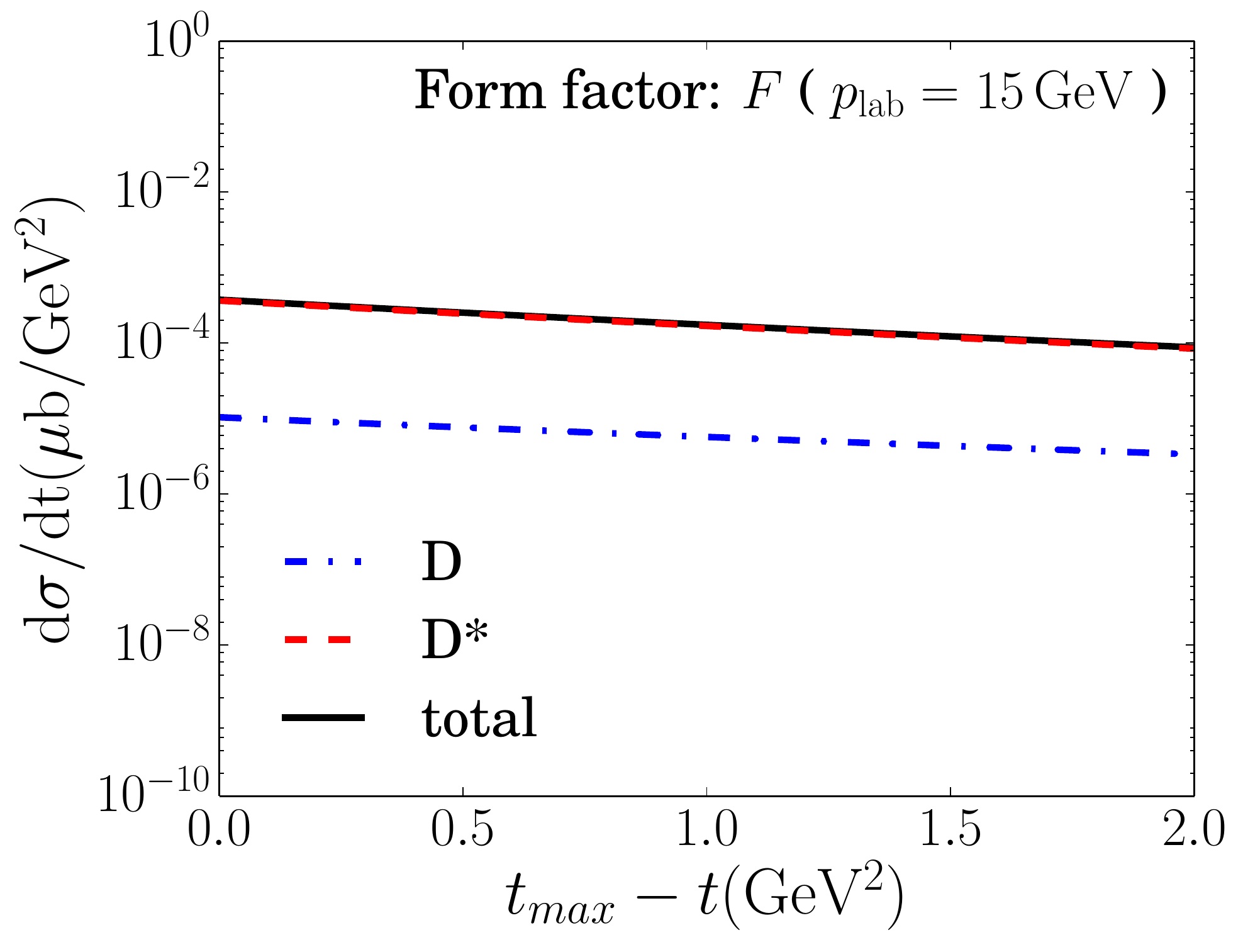}\includegraphics[width=5.5cm]{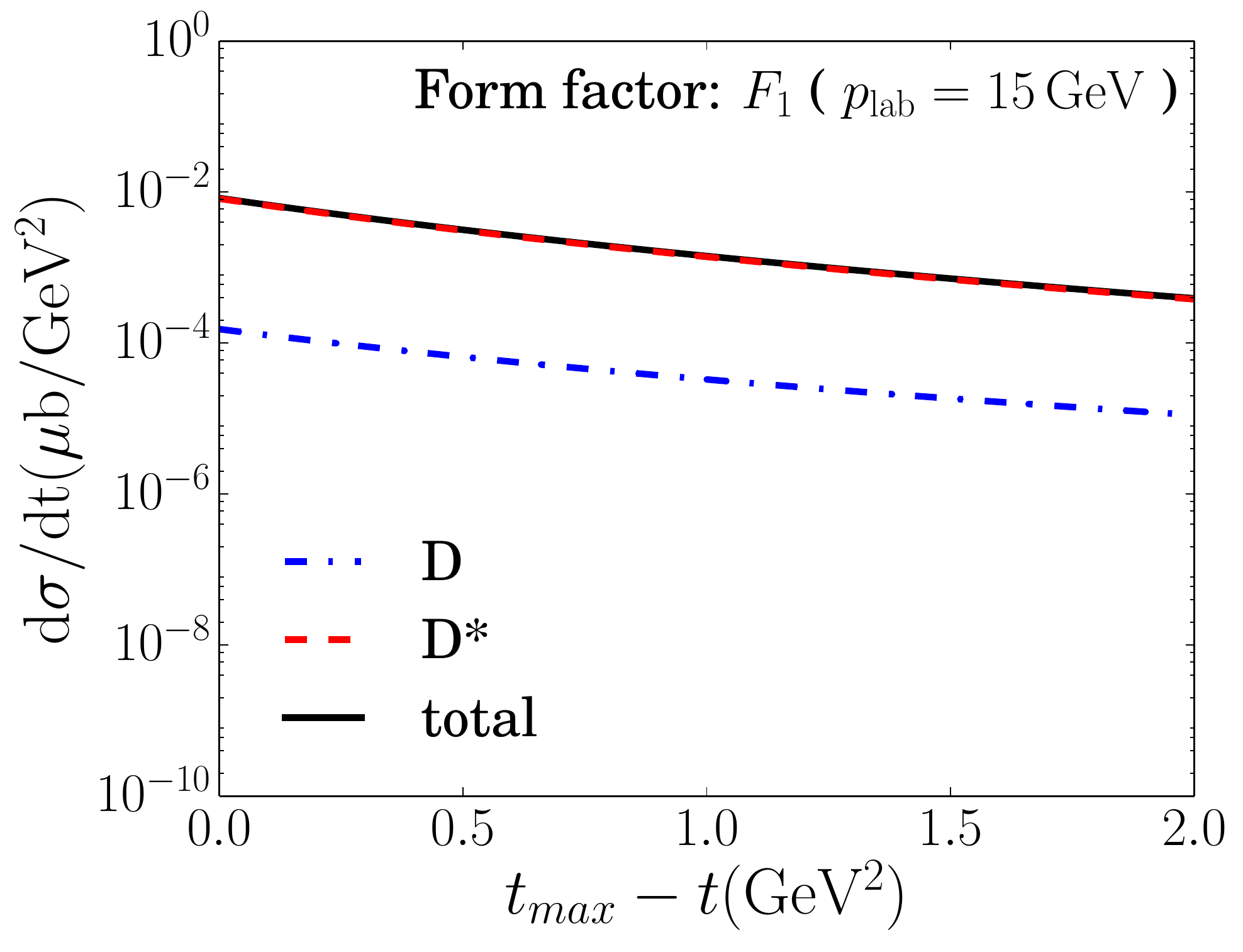}
\includegraphics[width=5.5cm]{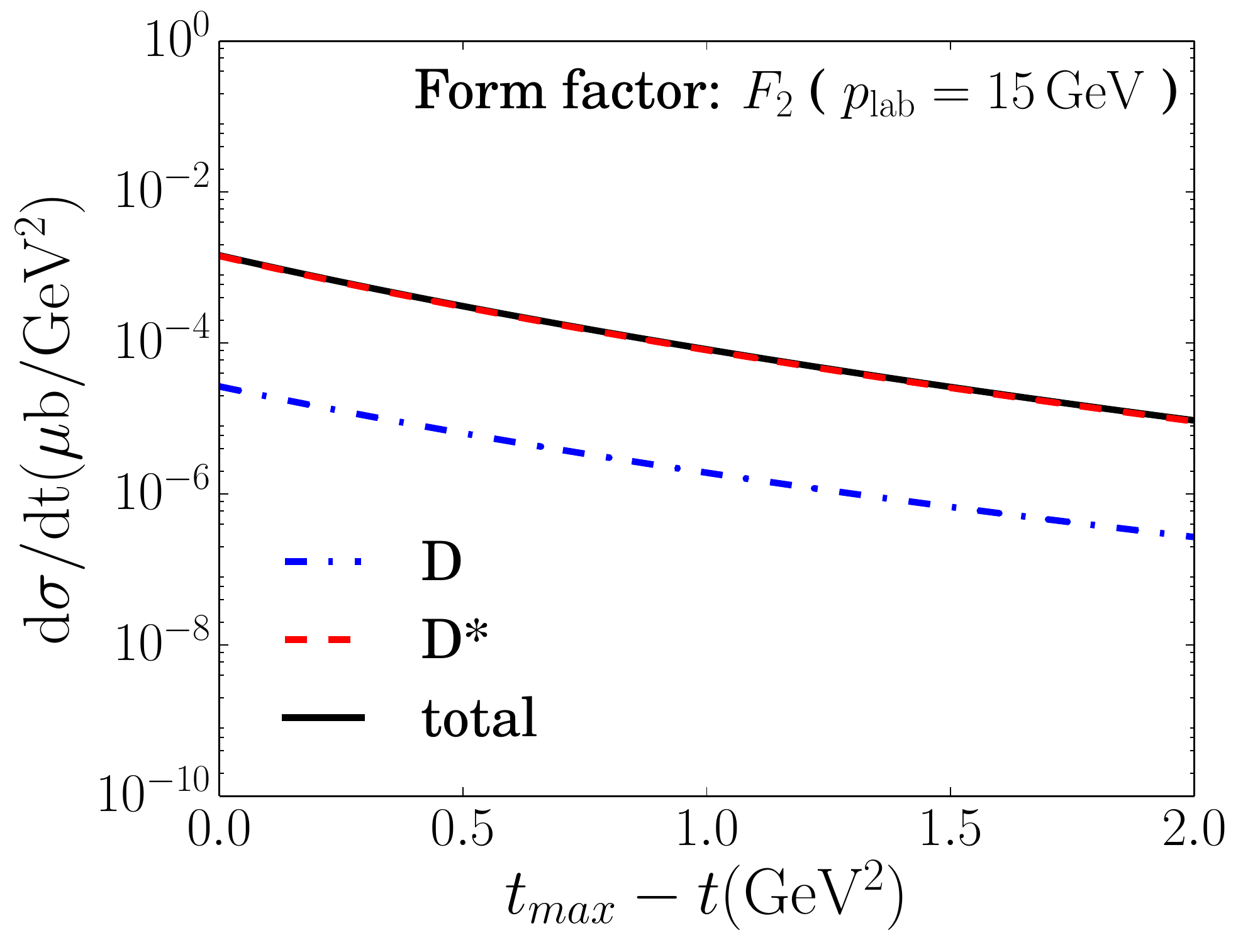}
\caption{Differential cross sections for $p\bar{p} \to \Sigma_{c}\bar{\Lambda}_{c}$ reaction at $p_{lab} = 15 \text{ GeV}$.}
\label{SLcEFT}
\end{figure}
\begin{figure}[H]
\centering
\includegraphics[width=5.5cm]{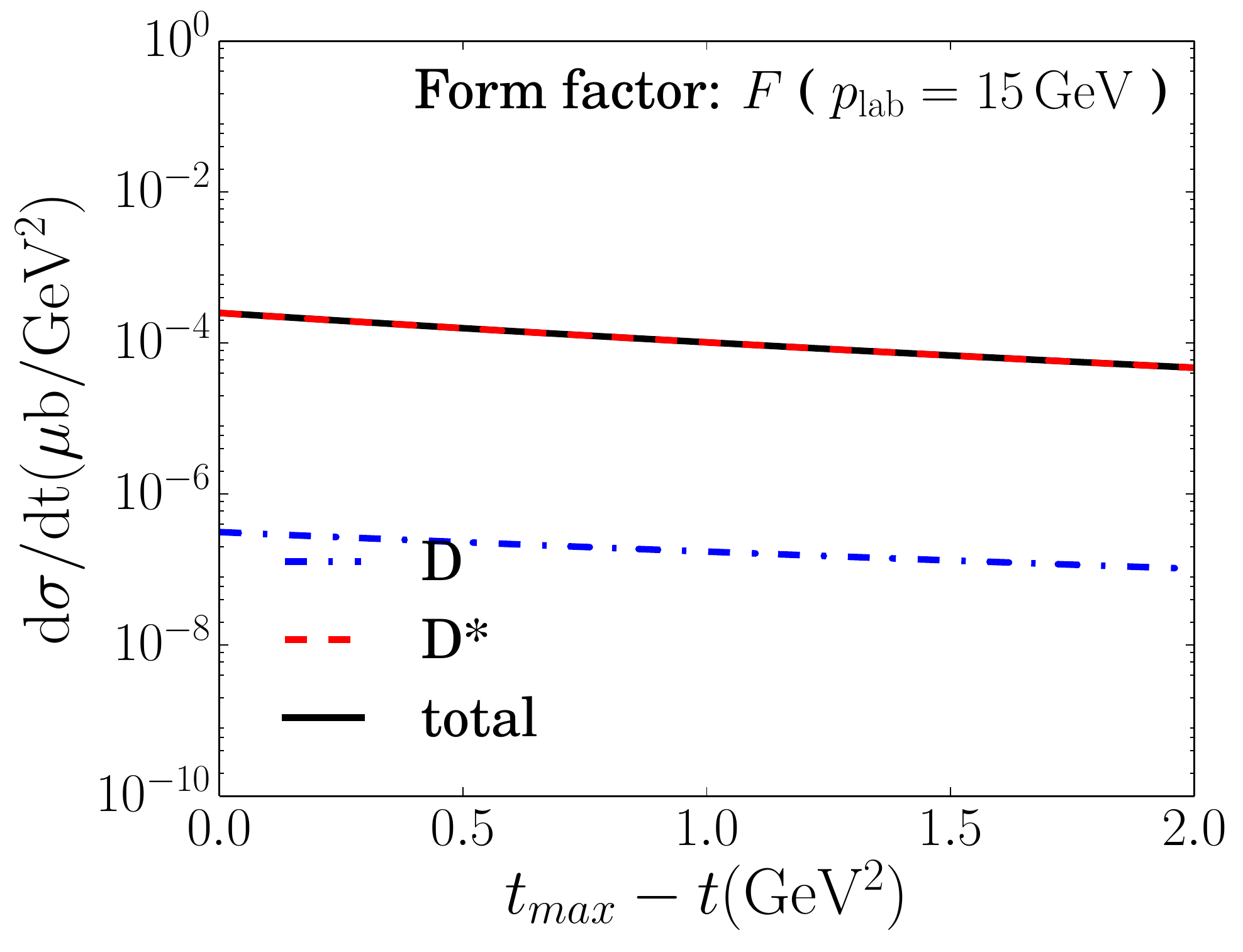}\includegraphics[width=5.5cm]{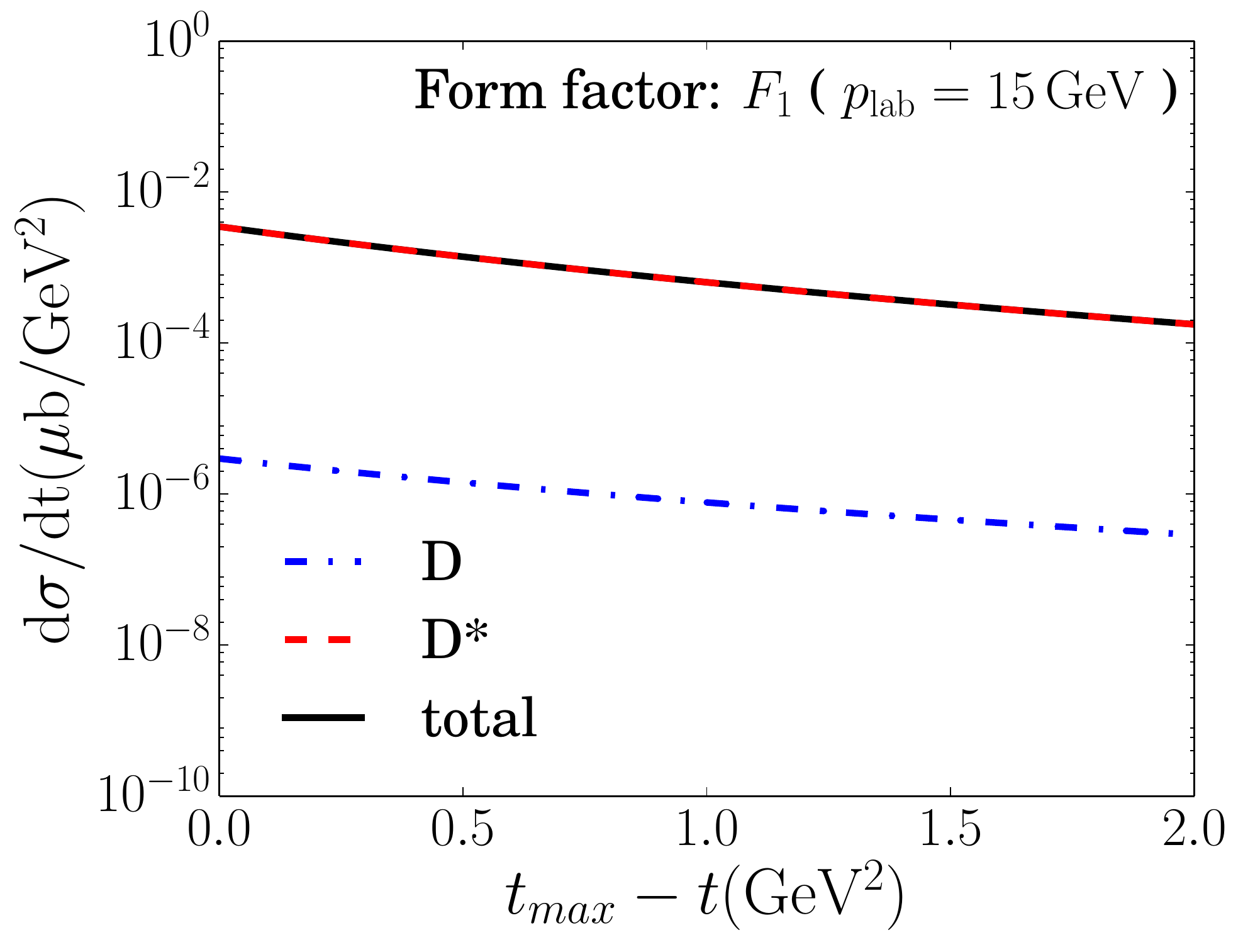}
\includegraphics[width=5.5cm]{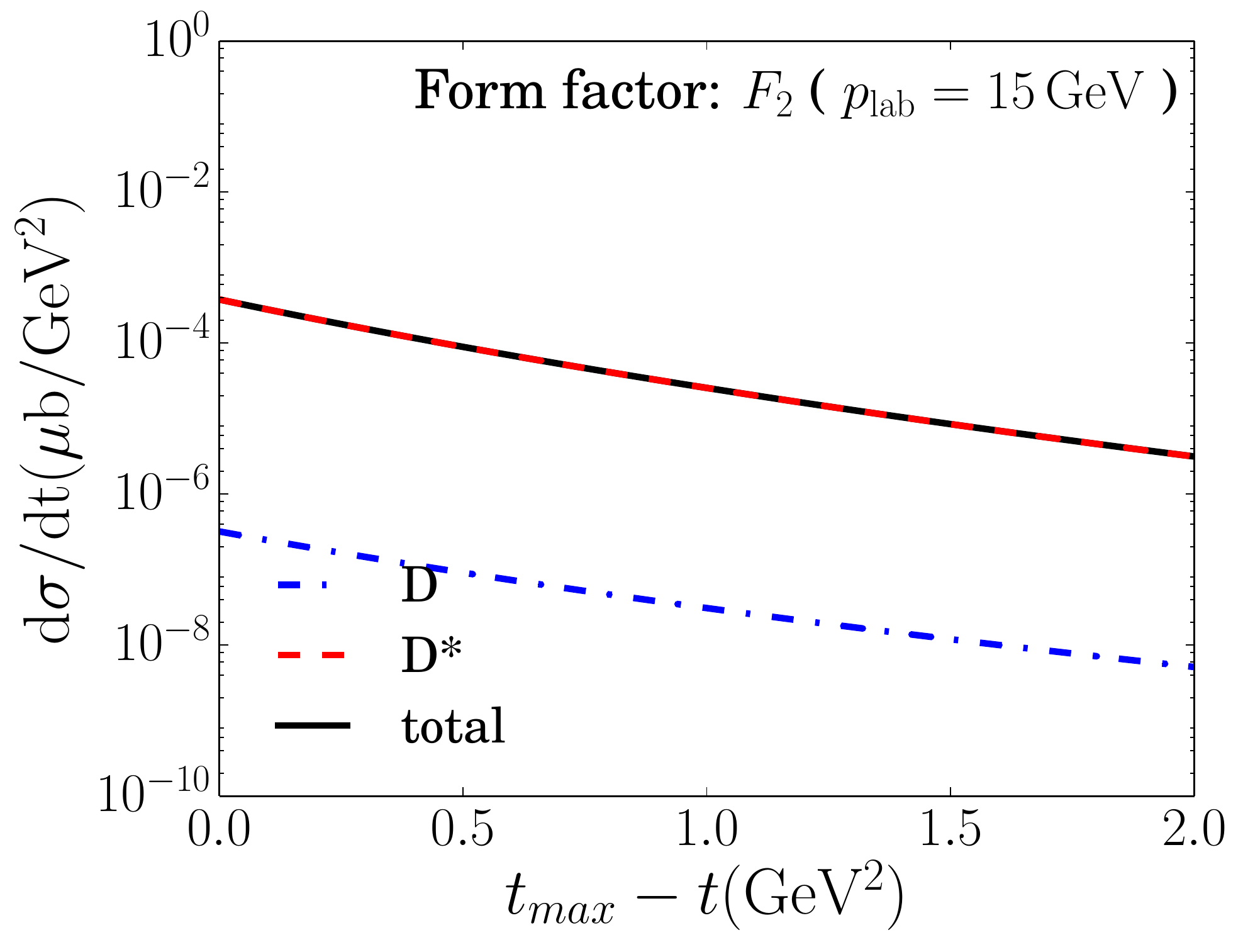}
\caption{Differential cross sections for $p\bar{p} \to \Sigma_{c}\bar{\Sigma}_{c}$ reaction at $p_{lab} = 15 \text{ GeV}$.}
\label{SScEFT}
\end{figure}

\indent \indent In this subsection, differential cross sections for charm productions are predicted by replacing strange hadrons by the charmed ones. 
The same set of coupling constants and cutoff parameters from strange sector is then applied to the charm production differential cross sections. 
The results for the reactions $p \bar{p} \to \Lambda_{c}\bar{\Lambda}_{c}$, $p \bar{p} \to \Sigma_{c}\bar{\Lambda}_{c}$, and $p \bar{p} \to \Sigma_{c}\bar{\Lambda}_{c}$ at $p_{lab} = 15 \text{ GeV}$ are presented in Fig.\ref{LLcEFT} to Fig.\ref{SScEFT}. 
Here we can see that $D^{*}$-meson exchange dominates differential cross sections and three different $t_{max}-t$ dependences are observed. 
The slope of the differential cross sections increases from the left to the right, which is similar to the observation with strangeness production cross sections. 
Near the forward angle region, the absolute values of the ones with the form factor $F_1$ and $F_2$ are almost at the same order, while the other ones with the form factor $F$ are much lower. 
If we consider the differential cross sections near forward angle production (i.e., $t_{max}-t = 0$), the suppression factor is in the order of $10^{-4}$ to $10^{-5}$. 
The highest suppression factor is provided by the form factor $F$, while the ones with the form factor $F_1$ and $F_2$ are lower. 
The absolute values of the differential cross sections as well as the contribution of $D$-meson exchange is decreased as the total mass of the final state is increased. 
From our calculations, differential cross sections with $(D + D^{*})$-meson exchange near the forward angle are in the order of $10^{-2}$ to $10^{-4}$ $\mu b /\text{GeV}^{2}$. 
Therefore, $t_{max} - t$ dependences as well as the suppression factors of charm production cross sections are mainly provided by the form factors. \newline 
\indent \indent Our predictions are in the same order as the results from Regge-based model in Ref.\cite{Titov:2008yf} and QGSM-based model in Ref.\cite{Khodjamirian:2011sp}. 
Since the Regge theory is applicable to reproduce the differential cross sections at high energies, effective Lagrangian calculations should be perform in such a way that the suppression factor is in the same order as the one given by Regge theory. 
This implies the reason why the form factor $F$ is included to the whole amplitude instead of at each vertex. 
In the study of charm productions from pion-induced reactions in Ref.\cite{Kim:2015ita}, this similar treatment for the form factor $F$ was applied. 
Even the significance of this agreement may not be realized at this moment, we hope that it will be testified by the future experiments at $\bar{\text{P}}$ANDA. 
\section{Regge Approach}
\indent \indent In this section, we present differential cross sections for $p \bar{p} \to Y\bar{Y}^{\prime}$ reactions obtained from Regge approach. 
This model can be used to reproduce the forward and backward peaks in high energy scatterings, 
which is its unique feature over the effective Lagrangian method. 
Regge amplitudes for pseudoscalar and vector Reggeon exchanges are constructed, 
the $s$- and $t$-dependences are collected in Regge propagator, 
while the spin information is given by the Feynman amplitude. 
To remove additional $s$- and $t$- dependences from the Feynman amplitude which is required for the conservation of unitarity, the normalization factor introduced in Ref.\cite{Titov:2008yf} is employed. 
To express the finite size of hadrons, 
we introduce the overall residual factor to enhance $s$- and $t$- dependences of differential cross sections. 
The results for strange productions are presented, 
and the predictions for charm productions are then given by the same assumption in the previous section.
\label{sec-3}
\subsection{Regge amplitudes}
\indent \indent In this subsection, Regge amplitudes for strangeness productions are presented then we proceed to charm productions. 
To start with, the Regge amplitude for $K$ and $K^{*}$ Reggeon exchanges are written by,
\begin{align}
\mathcal{M}_{K_{R}} &= \mathcal{M}_{K}\left(t-m_{K}^{2}\right)P^{R}_{K}\left(s,t\right), \label{MKR}\\
\mathcal{M}_{K^{*}_{R}} &= \mathcal{M}_{K^{*}}\left(t-m_{K^{*}}^{2}\right)P^{R}_{K^{*}}\left(s,t\right),
\label{MKsR}
\end{align}
where $P^{R}_{K}\left(s,t\right)$ and $P^{R}_{K^{*}}\left(s,t\right)$ are Regge propagator for $K$ and $K^{*}$ Reggeons. 
Their explicit expressions are written by \cite{Kim:2015ita},
\begin{align}
P^{R}_{K}\left(s,t\right) &= \Gamma\left[-\alpha_{K}\left(t\right)\right]\alpha_{K}^{\prime}\left(\dfrac{s}{s_{K}}\right)^{\alpha_{K}\left(t\right)}, \\
P^{R}_{K^{*}}\left(s,t\right) &= \Gamma\left[1-\alpha_{K^{*}}\left(t\right)\right]\alpha_{K^{*}}^{\prime}\left(\dfrac{s}{s_{K^{*}}}\right)^{\alpha_{K^{*}}\left(t\right)-1}.
\label{Regge Propagator K}
\end{align}
In other words, Regge amplitude is obtained by replacing the Feynman propagator with the Regge one. 
Here, the information of the ground states and the series of excited states lies on the same trajectory is collected in the function $\alpha\left(t\right)$, namely ``Regge trajectory''. 
Unlike effective Lagrangian method, single particle exchange is now replaced by an exchange of the family of particles in the $t$-channel exchange processes. 
Regge trajectories for $K$ and $K^{*}$ Reggeons taken from Ref.\cite{Brisudova:1999ut} are
\begin{align}
\alpha_{K}\left(t\right) &= -0.15 + 0.62t, \\
\alpha_{K^{*}}\left(t\right) &= 0.41+0.71t.
\end{align}
The scaling parameters $s_{K} = 2.42 \text{ GeV}^{2}$ and $s_{K^{*}} = 2.45 \text{ GeV}^{2}$ are determined from QGSM \cite{Kaidalov:1994mda}. 
At high energies, the imaginary part of the inelastic scattering amplitude is factorized into a product between two elastic scattering amplitudes. 
This factorization provides the useful prescription to evaluate the scaling and Regge parameters for inelastic scatterings. 
It is important to emphasize that, the $\Gamma$-function in the Regge propagator resembles the behavior of the Feynman propagator near the pole of the ground state lies on that Regge trajectory. 
For example, If we expand the $\Gamma$-function as a function of $\alpha_{K^{*}}\left(t\right)$ near the point $t = m^{2}_{K^{*}}$ where $\alpha\left(t=m^{2}_{K^{*}}\right) = 1$, the following behavior,
\begin{equation}
\Gamma\left(1-\alpha_{K^{*}}\left(t\right)\right) \simeq -\dfrac{1}{\alpha_{K^{*}}^{\prime}\left(t-m_{K^{*}}^{2}\right)},
\end{equation}
is clearly observed. The following overall residual factor which expresses the finite structure of hadrons,
\begin{equation}
C\left(t\right) = \dfrac{k}{\left(1-\dfrac{t}{\Lambda^{2}}\right)^{2}},
\end{equation}
is introduced to improve $s$- and $t$- dependences in Regge amplitudes. 
There are two unknown parameters in this residual factor: normalization constant $k$ and cutoff parameter $\Lambda$. 
These parameters will be fixed by comparing differential cross sections based on $(K+K^{*})$ Reggeon exchanges to the experimental data for strangeness productions. 
The total Regge amplitude of the reaction $p\bar{p} \to Y\bar{Y}^{\prime}$ is written by
\begin{equation}
\mathcal{M}^{R}_{p\bar{p} \to Y\bar{Y}^{\prime}} = C\left(t\right)\left(\mathcal{M}^{R}_{K}+\mathcal{M}^{R}_{K^{*}}\right).
\label{ReggeM}
\end{equation}
In case of the reaction $p\bar{p} \to Y_{c}\bar{Y}_{c}^{\prime}$, Regge propagators and scaling parameters for $D$ and $D^{*}$ Reggeons are \cite{Brisudova:1999ut},
\begin{align}
P^{R}_{D}\left(s,t\right) &= \Gamma\left[-\alpha_{D}\left(t\right)\right]\alpha_{D}^{\prime}\left(\dfrac{s}{s_{D}}\right)^{\alpha_{D}\left(t\right)}, \label{RPD}\\
P^{R}_{D^{*}}\left(s,t\right) &= \Gamma\left[1-\alpha_{D^{*}}\left(t\right)\right]\alpha_{D^{*}}^{\prime}\left(\dfrac{s}{s_{D^{*}}}\right)^{\alpha_{D^{*}}\left(t\right)-1}, \label{RPDs}\\
\alpha_{D}\left(t\right) &= -1.61 + 0.44t, \label{Alpha D}\\
\alpha_{D^{*}}\left(t\right) &= -1.02+0.47t.
\label{Alpha Ds}
\end{align}
In Eq.(\ref{RPD}) and Eq.(\ref{RPDs}), the scaling parameters for $D$ and $D^{*}$ Reggeon exchanges are $S_{D} = 5.46 \text{ GeV}^{2}$ and $S_{D^{*}} = 6.01 \text{ GeV}^{2}$. Regge formalism is consistent with unitarity, it can be used to reproduce high energy scattering cross sections, especially in the forward angle region. As $s \to \infty$, the following asymptotic behavior of differential cross sections,
\begin{equation}
\dfrac{d\sigma}{dt}\left(s \to \infty, t \to 0 \right) \propto s^{2\left(\alpha(t)-1\right)},
\label{asympt}
\end{equation}
is satisfied. 
\subsection{Normalization of Regge amplitudes}
\indent \indent As we have seen in the explicit form of Regge amplitudes in Eq.(\ref{MKR}), the $s$- and $t$- dependences appear in the Feynman amplitude as well as Regge propagator and residual factor. In Regge theory, these dependences should come from the Regge propagator only, otherwise the conservation of unitarity will be violated. Therefore, the following normalization factor is introduced in Ref.\cite{Titov:2008yf} to remove additional $s$- and $t$- dependences from the Feynman amplitude,
\begin{equation}
\mathcal{N}\left(s,t\right) = \dfrac{A^{\infty}\left(s\right)}{A\left(s,t\right)}, \hspace{1 cm} A^{2}\left(s,t\right) = \sum_{s_{3},s_{4}} \left|M\left(s,t\right)\right|^{2}.
\label{normalization}
\end{equation}
The function $M(s,t)$ is gotten by removing the denominator of the Feynman propagator from the Feynman amplitude $\mathcal{M}$. The term $A^{\infty}\left(s\right)$ is the leading term in $M(s,t)$ at large s. For pseudoscalar meson exchange, we get $A^{\infty}\left(s\right) = 2\left(m_{N}-m_{Y}\right)\left(m_{\bar{N}}-m_{\bar{Y}^{\prime}}\right)$ while $A^{\infty}\left(s\right) = 4s$ for vector meson exchange. Therefore, Regge amplitude in Eq.(\ref{ReggeM}) is then rewritten by,
\begin{equation}
\mathcal{M}^{R}_{p\bar{p} \to Y\bar{Y}^{\prime}} = C\left(t\right)\left(\mathcal{N}_{K}\left(s,t\right)\mathcal{M}^{R}_{K}+\mathcal{N}_{K^{*}}\left(s,t\right)\mathcal{M}^{R}_{K^{*}}\right).
\label{ReggeN}
\end{equation}
This form of Regge amplitude is then consistent with the conservation of unitarity.
\subsection{Results for Strangeness Productions}
\begin{figure}[H]
\centering
\includegraphics[width=5.5cm]{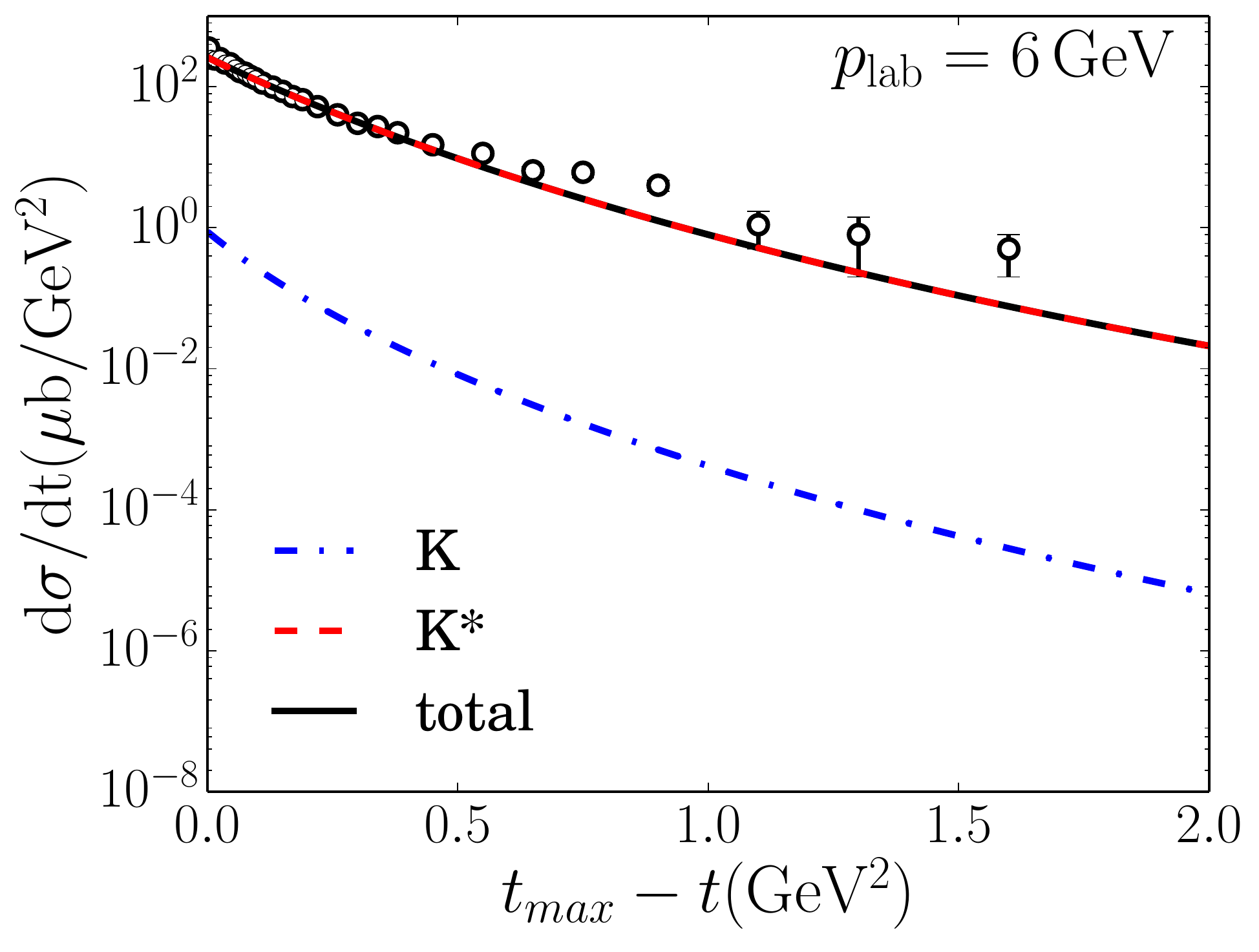}
\includegraphics[width=5.5cm]{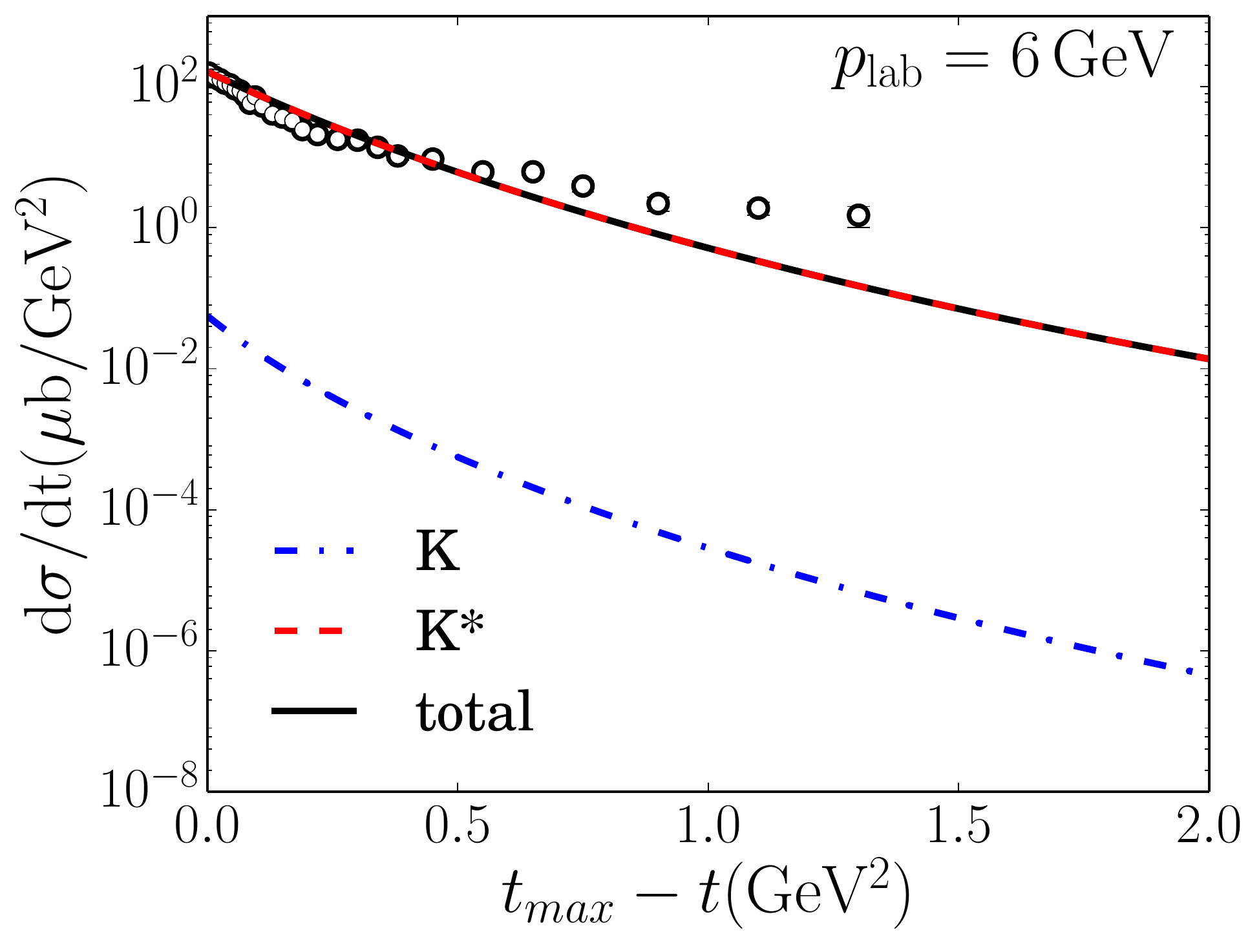}
\includegraphics[width=5.5cm]{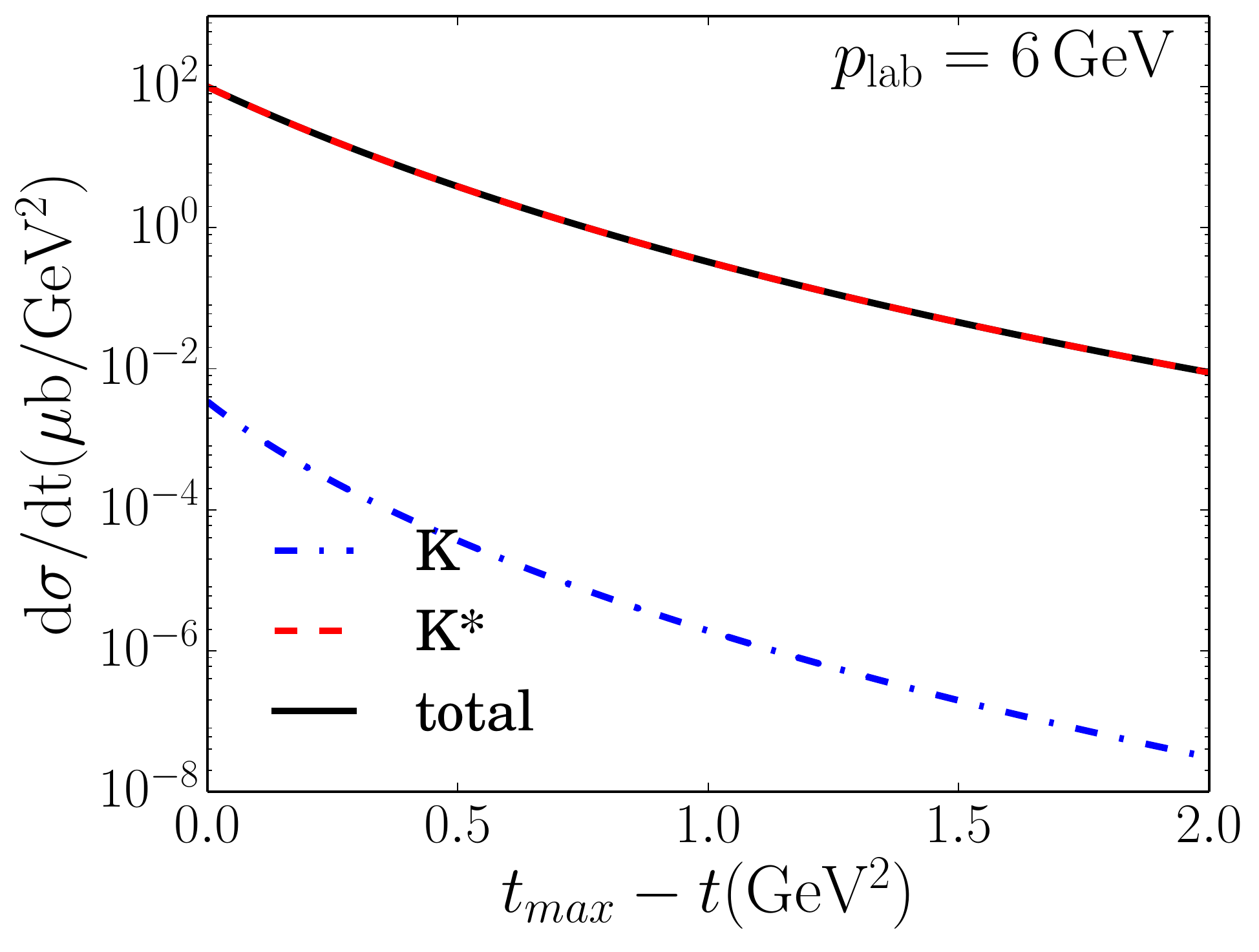}
\caption{Differential cross sections for the reaction (left)$p\bar{p} \to \Lambda\bar{\Lambda}$, (middle)$p\bar{p} \to \Sigma\bar{\Lambda}$, (right)$p\bar{p} \to \Sigma\bar{\Sigma}$ at $p_{lab} = 6 \text{ GeV}$. The circles correspond to the experimental data taken from Ref.\cite{Becker:1978kk}.}
\label{LBR}
\end{figure}
\indent \indent In this subsection, differential cross sections for strangeness productions are computed in Regge approach. 
The results with the same set of coupling constants from effective Lagrangian method are presented in Fig.\ref{LBR}. 
Similar to effective Lagrangian method, we need to fix the unknown parameters by comparing differential cross sections with $(K+K^{*})$ Reggeon exchange to the observed differential cross sections of the reactions $p\bar{p}\to\Lambda\bar{\Lambda}$ and $p\bar{p}\to\Sigma\bar{\Lambda}$. 
Following parameters for the overall residual factor,
\begin{equation}
C\left(t\right): \hspace{0.2cm} k = 0.25, \hspace{0.2cm} \Lambda = 1.1 \text{  GeV},
\end{equation}
are fixed such that the differential cross sections in Fig.\ref{LBR} agree with the experimental data. 
This set of parameters is then applied to all production reactions under consideration. 
As mentioned earlier, differential cross sections near the forward angle production can be well reproduced in Regge approach. 
Therefore, the experimental data at the low $t_{max}-t$ region is of interested. 
In Fig.\ref{LBR}, one can observe that, differential cross sections for strangeness productions with $(K + K^{*})$ Reggeon exchange are in a good agreement with the data compared to the ones from effective Lagrangian method. 
Especially, the slope of the data can be reproduced well even the absolute values are underestimated than the data in the finite angle region. 
The differential cross section of the reaction $p\bar{p}\to\Lambda\bar{\Lambda}$ with $(K + K^{*})$ Reggeon exchange is in the same order as the one with the form factor $F_{2}$. In case of the reaction $p\bar{p}\to\Sigma\bar{\Lambda}$, the differential cross section is similar to the one with the form factor $F_{1}$. 
Similar to the results from effective Lagrangian method, differential cross sections with $K$ Reggeon exchange are largely suppressed by the factor $10^{-2}$ to $10^{-5}$, while the ones with $K^{*}$ Reggeon exchange almost overlap with the results for $(K + K^{*})$ Reggeon exchange. 
However, differential cross sections with $K$ Reggeon exchange decrease faster than the ones from effective Lagrangian method, 
which indicates the important role of $K^{*}$ Reggeon exchange at this energy.
\subsection{Results for Charm Productions}
\begin{figure}[H]
\centering
\includegraphics[width=5.5cm]{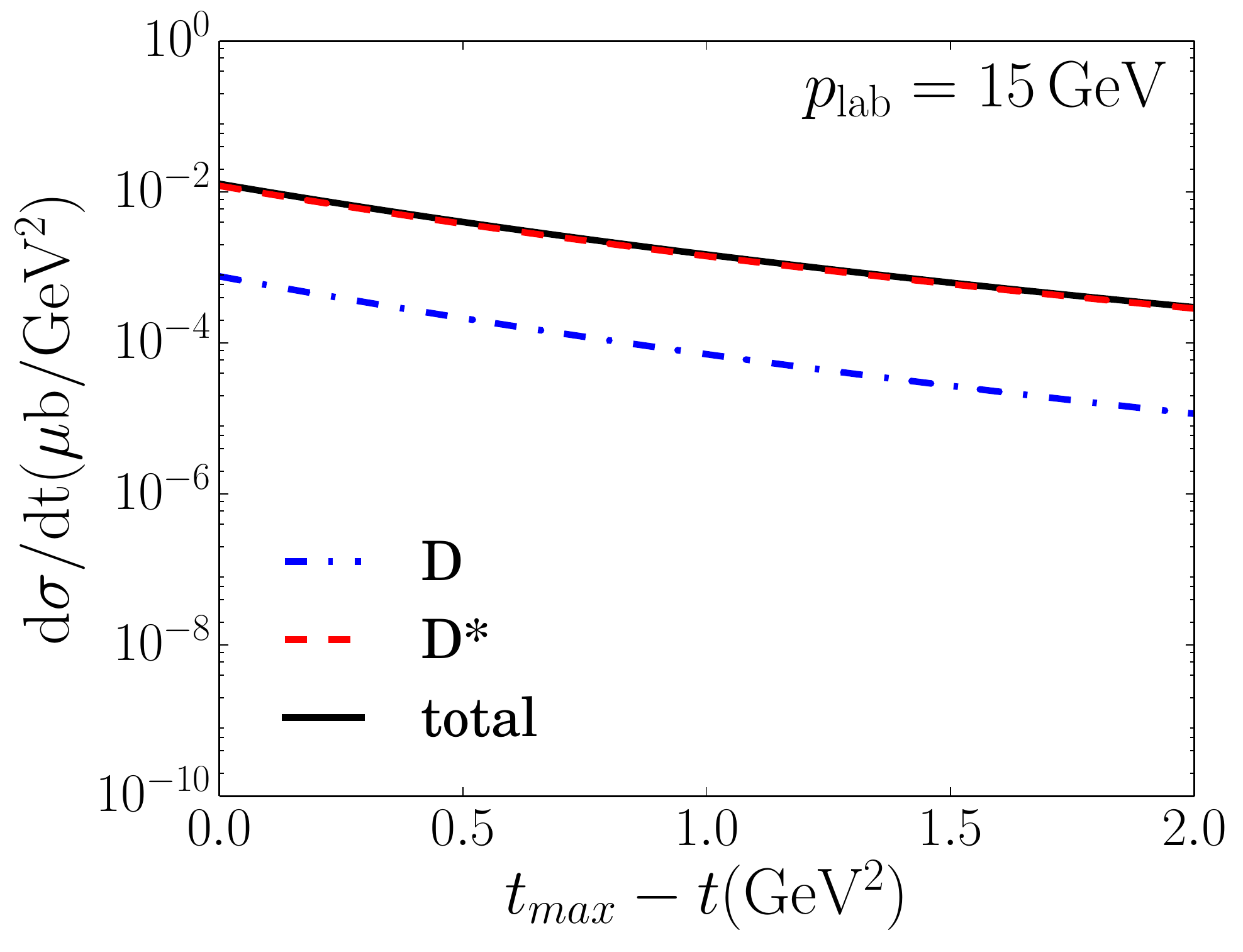}
\includegraphics[width=5.5cm]{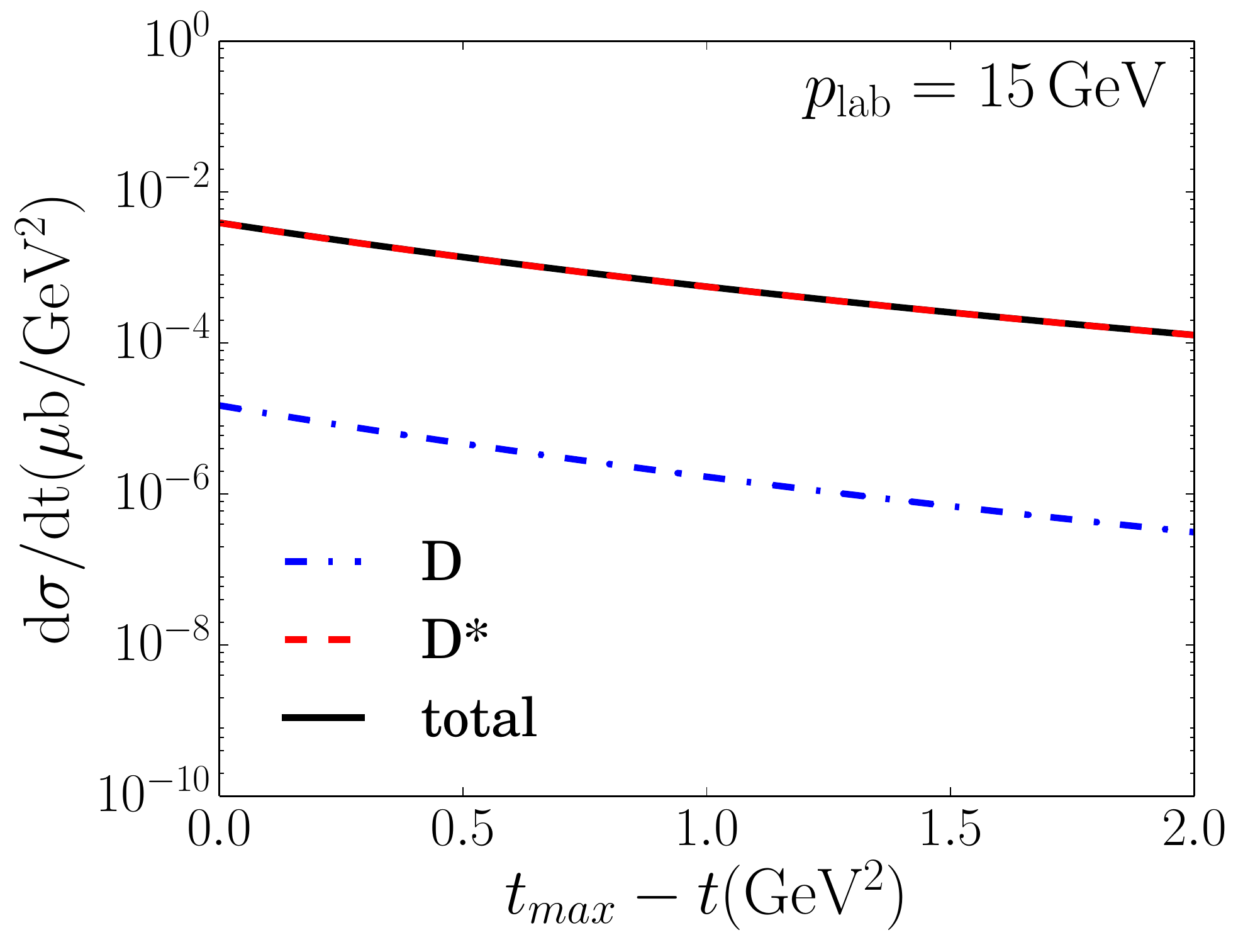}
\includegraphics[width=5.5cm]{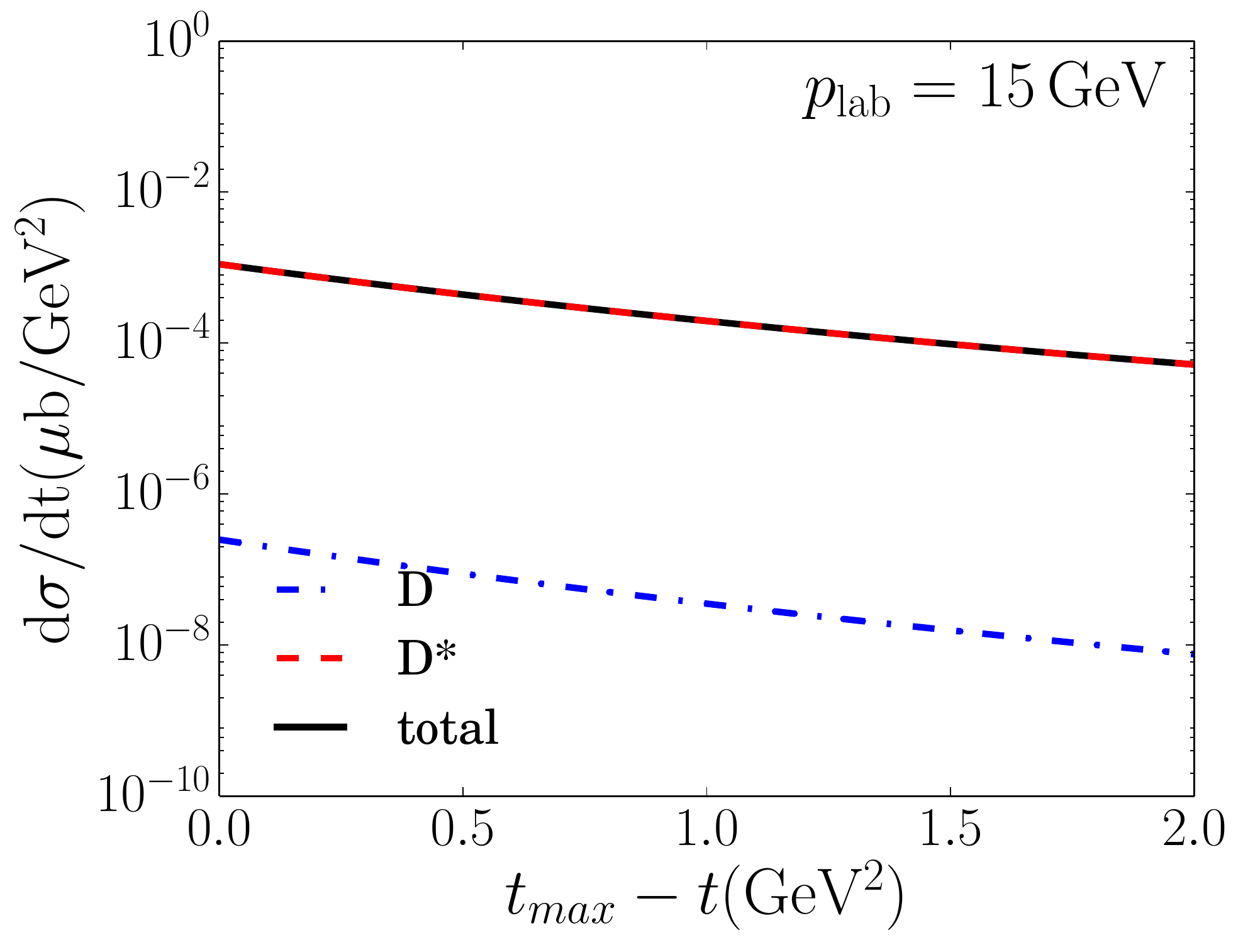}
\caption{Differential cross sections for the reactions $p\bar{p} \to \Lambda_{c}\bar{\Lambda}_{c}$(left), $p\bar{p} \to \Sigma_{c}\bar{\Lambda}_{c}$(middle), $p\bar{p} \to \Sigma_{c}\bar{\Sigma}_{c}$(right) at $p_{lab} = 15 \text{ GeV}$.}
\label{HBR}
\end{figure}

\indent \indent By replacing strange hadrons by charm hadrons, 
differential cross sections for charm productions are given. 
The results are collected in Fig.\ref{HBR}. 
Obviously, they are similar to the ones with the form factor $F_{2}$. 
The suppression factors in the order of $10^{-4}$ to $10^{-5}$ from the strangeness production cross sections are observed, 
which agree with the other ones from effective Lagrangian approach. 
Here, the absolute values of differential cross sections with $D$ Reggeon exchange become less important as the total mass of the outgoing charmed baryons is increasing. 
Similar to strangeness productions, the differential cross sections contributed by $D$ Reggeon are largely suppressed, while the ones with $D^{*}$ Reggeon exchange are more important. 
The absolute values of differential cross sections from our predictions is in the order of $10^{-2}$ to $10^{-4} \mu b / \text{GeV}^{2}$ near the forward angle region, which agrees with the results reported in Ref.\cite{Titov:2008yf, Khodjamirian:2011sp}.
\section{Comparison of the two models}
\label{sec-4}
\begin{figure}[H]
\centering
\includegraphics[width=5.5cm]{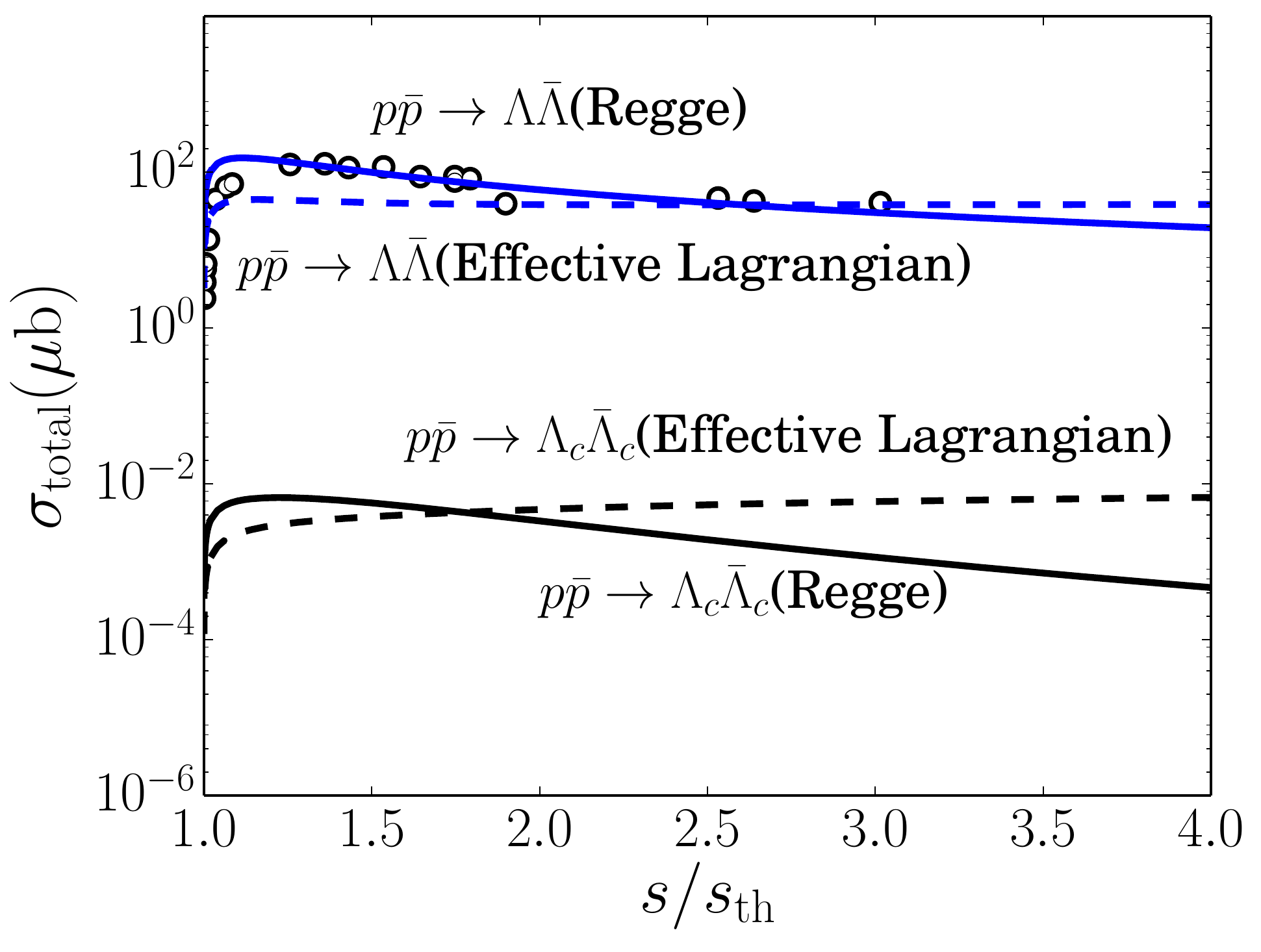}
\includegraphics[width=5.5cm]{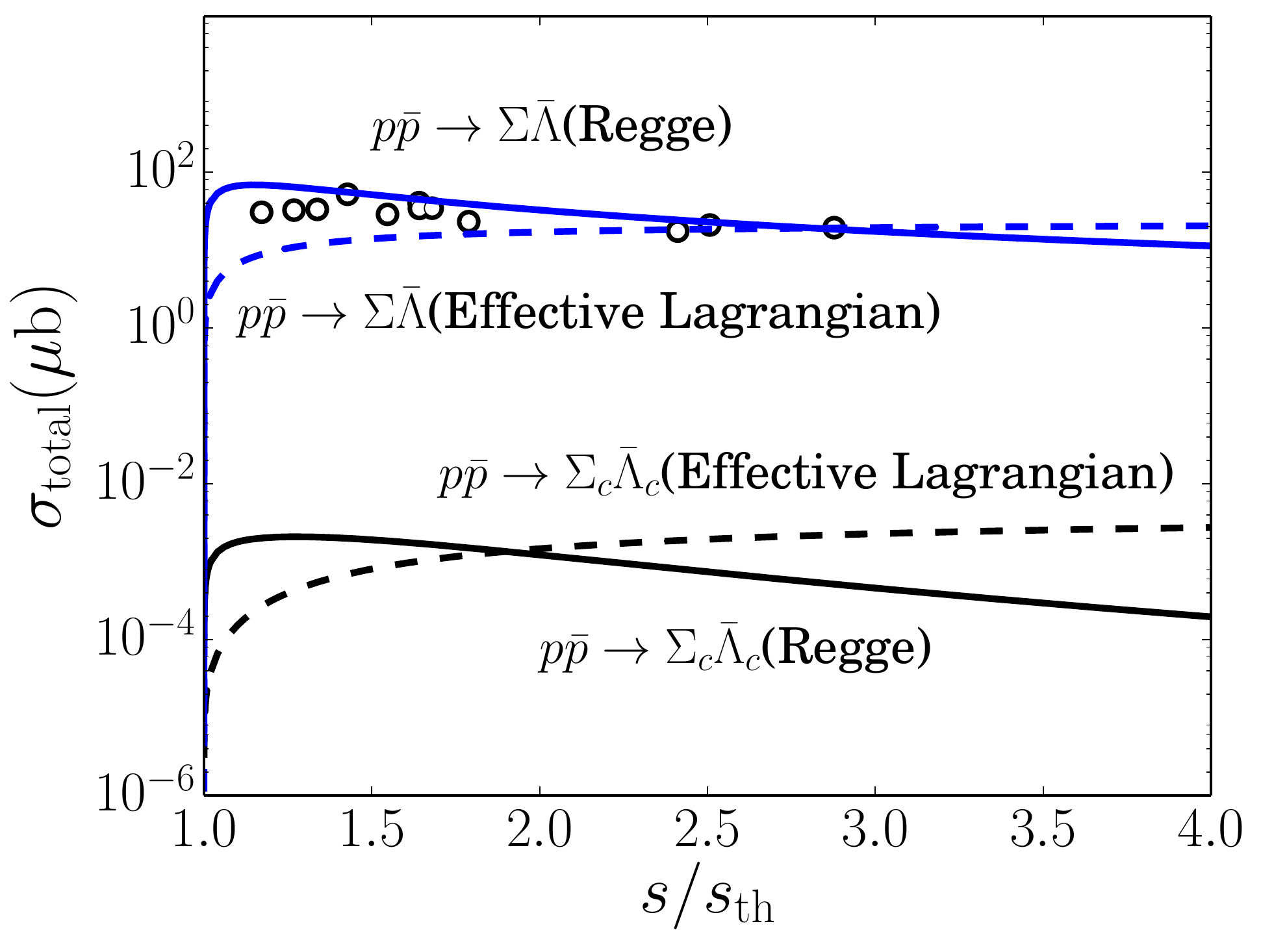}
\includegraphics[width=5.5cm]{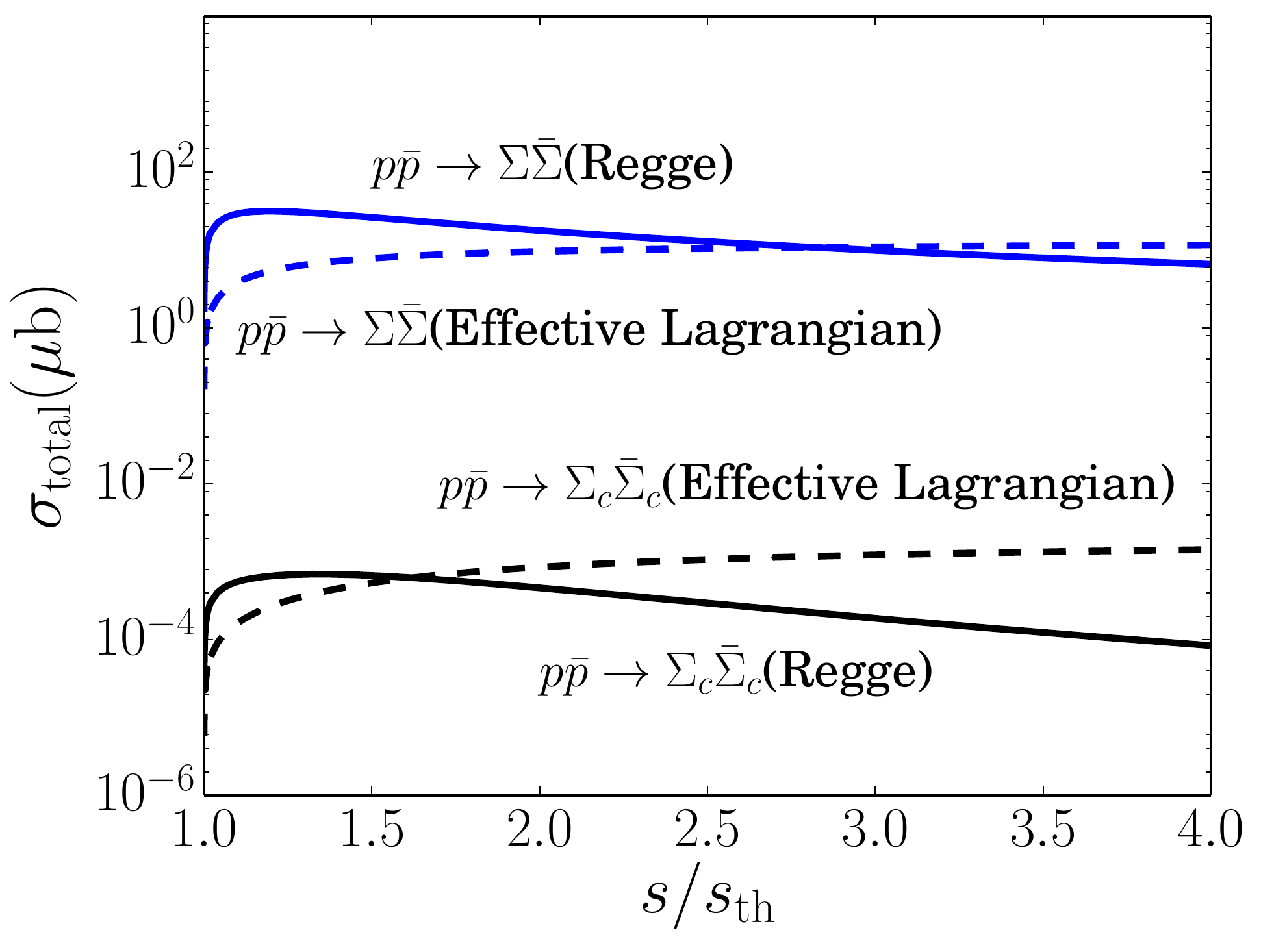}
\caption{Total cross sections for the reactions (left) $p\bar{p}\to\Lambda\bar{\Lambda}$ and $p\bar{p}\to\Lambda_{c}\bar{\Lambda}_{c}$ (middle) $p\bar{p}\to\Sigma\bar{\Lambda}$ and $p\bar{p}\to\Sigma_{c}\bar{\Lambda}_{c}$. (right) $p\bar{p}\to\Sigma\bar{\Sigma}$ and $p\bar{p}\to\Sigma_{c}\bar{\Sigma}_{c}$. The circles are experimental data taken from Ref.\cite{Becker:1978kk,Barnes:1989je,Barnes:1990bs,Barnes:1996si}.}
\label{tc}
\end{figure}
\indent \indent In this section, total cross sections for strangeness and charm productions from effective Lagrangian and Regge approaches are presented. 
This can be done by performing an integration for ${d\sigma}/{dt}$ over $t$. 
Then, total cross sections are displayed as functions of $s/s_{th}$, where $s_{th} = \left(m_{Y}+m_{\bar{Y}^{\prime}}\right)^{2}$. 
Total cross sections for various strangeness and charm productions are displayed in Fig.\ref{tc}. 
For effective Lagrangian results, we display the total-cross sections with the form factor $F$. 
Since we have shown the strange and charm productions in the same figure, we need to emphasize that their $s/s_{th}$ are different. \newline
\indent \indent We start with the total cross sections for the reactions $p\bar{p}\to\Lambda\bar{\Lambda}$ and $p\bar{p}\to\Lambda_{c}\bar{\Lambda}_{c}$ in the left panel of Fig.\ref{tc}. 
Here, the total cross sections of the reactions $p\bar{p}\to\Lambda\bar{\Lambda}$ from Regge and effective Lagrangian approaches agree in the narrow region near the point $s/s_{th} \sim 2.7$. This is due to the agreements between differential cross sections from both methods and the experimental data at $p_{lab} = 6 \text{ GeV}$. 
As we go from this point to the near threshold region (i.e., $s/s_{th} \in (1,1.09]$), the total cross section from effective Lagrangian approach can reproduce the data well while the other one from Regge approach is overestimated than the experimental data. 
In contrast, as we go to the high energy region (i.e., $s/s_{th} \geq 1.2$), the one from Regge approach seems to be in a better agreement with the data in this region. 
However, the effective Lagrangian result underestimates for the first eight data points. 
After the point where $s/s_{th} \approx 1.8$ is reached, total cross section agrees better with the data. 
The asymptotic behavior of the total cross section from Regge approach is decreasing while the effective Lagrangian result seems to diverge at high $s/s_{th}$.
Total cross sections for the reaction $p\bar{p}\to\Lambda_{c}\bar{\Lambda}_{c}$ are that of the bottom part of the plot. 
Near the threshold, the $\Lambda_{c}\bar{\Lambda}_{c}$ production cross section is suppressed by the factor $10^{-4}$ for effective Lagrangian approach and $10^{-5}$ for Regge approach. Their $s/s_{th}$ dependences are similar to $\Lambda\bar{\Lambda}$ production cross sections. 
Near the production threshold, our predictions are in the same order as the results from Ref.\cite{Kroll:1988cd,Goritschnig:2009sq,Kaidalov:1994mda}. The predictions from Ref.\cite{Haidenbauer:2016pva,Shyam:2014dia} seems to be overestimated compared to our results.
\newline
\indent \indent For the reactions $p\bar{p}\to\Sigma\bar{\Lambda}$ and $p\bar{p}\to\Sigma_{c}\bar{\Lambda}_{c}$, total cross sections are displayed in the middle panel of Fig.\ref{tc}. 
The total cross sections for the reaction $p\bar{p}\to\Sigma\bar{\Lambda}$ from both methods agree with each other in the narrow region near the point $s/s_{th} \sim 2.5$. 
As we interpolate to the lower energy region, the one from Regge method seems to be overestimated with the data compared to other one from effective Lagrangian method which is underestimated. 
Total cross sections from both methods agree with the data in the range where $s/s_{th} \in [2.4,2.9]$.
The asymptotic behavior is observed from the Regge result, while the one from effective Lagrangian approach is increased at high $s/s_{th}$. 
In this case, the total cross sections in charm sector gain a suppression factor $10^{-4}$ for both methods. 
Our results is underestimated as compared to the results in Ref.\cite{Shyam:2014dia}.
\newline
\indent \indent For the reactions $p\bar{p}\to\Sigma\bar{\Sigma}$ and $p\bar{p}\to\Sigma_{c}\bar{\Sigma}_{c}$, total cross sections are predicted in the right panel of Fig.\ref{tc}. 
In this case, the experimental data for the reaction $p\bar{p}\to\Sigma\bar{\Sigma}$ is not available. 
The suppression factors are the same as the one between the reactions $p\bar{p}\to\Sigma\bar{\Lambda}$ and $p\bar{p}\to\Sigma_{c}\bar{\Lambda}_{c}$. 
Again, our results is smaller as compared to the results in Ref.\cite{Shyam:2014dia}.
\newline
\indent \indent
In Ref.\cite{Haidenbauer:1991kt},  different cutoff parameter for $KN\Lambda$ and $K^{*}N\Lambda$ vertices is employed for the total cross sections of $p\bar{p} \to \Lambda\bar{\Lambda}$ near the threshold, $\Lambda_{KN\Lambda} = \Lambda_{K^{*}N\Lambda} = 1.2$ GeV.   In our present model, we can employ these larger cutoff parameters.  In this case, however,  we need to reduce the strengths of the coupling constants, 
in particular the dominant one for the $K^{*}$ meson, to reproduce the observed total cross sections.  
However, in this case the $t-$ dependence of ${d\sigma}/{dt}$ at $p_{L} = 6 \text{ GeV}$ cannot be reproduced well.
Furthermore, when extending their calculations to the charm sector~\cite{Haidenbauer:2016pva}, a larger cutoff parameter of 3 GeV was employed.  
As a result they have a significantly larger production rates for the charmed baryon productions, typically $10^{-1}$ times strangeness production rates.  
If we consider the physical meaning of the cutoff as related to the hadron size $\sim$ 1 fm, it is reasonable to employ a cutoff of order 1 GeV  that leads to the suppression of order $10^{-4} - 10^{-5}$.
\section{Summary and Conclusion}
\indent \indent We have presented the study of strangeness and charm productions in effective Lagrangian and Regge approaches. 
The relevant effective Lagrangians are employed to compute differential cross sections of strangeness productions in the diffractive region where $t$-channel dynamics is dominant. 
The coupling constants for $K$ meson are determined by $SU(3)$ relations while other unknown parameters for $K^{*}$ couplings and for form factors are fixed by existing observed data for strangeness productions. 
To express the finite size of hadrons, three different form factors are employed. 
The $t$-dependences and absolute values of the observed strangeness production cross sections are reproduced by the differential cross sections based on $(K+K^{*})$-meson exchanges if the appropriate set of parameters is fixed. 
Differential cross sections for charm productions are given by replacing strange hadrons by the charmed ones, different suppression factors are observed in the order of $10^{-4}$ to $10^{-5}$. 
The absolute values of charm production cross sections from $10^{-2}$ to $10^{-4}$ $\mu b /\text{GeV}^{2}$ are observed. \newline
\indent \indent Then we proceed to the Regge approach by replacing Feynman propagator by Regge ones. 
The scaling parameters are obtained from QGSM prescription. 
Parameters for the overall residual factor is fixed by comparing differential cross sections with $(K+K^{*})$ Reggeon exchanges to the observed strangeness production cross sections. 
The additional normalization factor is also introduced to remove additional $s$- and $t$-dependence from the Feynman amplitudes. 
The results obtained from Regge theory are similar to the ones obtained from effective Lagrangian with the form factor $F_{2}$, the $t$- dependences of the data near the forward angle region are well reproduced compared to the results from effective Lagrangian approach. 
Then, differential cross sections for charm productions are given by replacing strange by charm hadrons. The suppression factors and the absolute values of the differential cross sections agree with the ones obtained from effective Lagrangian approach.\newline
\indent \indent Finally, we extend the results to other energy regions by performing an integration over $t$ to the differential cross sections. 
The total cross sections for strange and charm productions are obtained. 
Experimental cross section for $\Lambda\bar{\Lambda}$ productions in the low energy region can be well explained by effective Lagrangian result, while the high energy behavior can be well explained by Regge result. In case of $\Sigma\bar{\Lambda}$ productions, both effective Lagrangian and Regge results can reproduce the data in the high energy region while the low energy behavior cannot be reproduced quite well.   
From our models, the charm production cross sections are suppressed by the factor from $10^{-4}$ to $10^{-5}$ compared to the ones for strangeness productions. 
Our study predicts the total cross sections for charm productions in the order of $10^{-2}$ to $10^{-4}$ $\mu b$ near threshold. This prediction will be testified by the future $\bar{\text{P}}$ANDA experiment.
\label{sec-5}
\section{Acknowledgement}
\indent \indent This work is supported by Thailand Science Research and Innovation, Suranaree University of Technology, Center of Excellence in High Energy Physics and Astrophysics, Suranaree University of Technology, Research Center for Nuclear Physics, Osaka University, and a FrontierLab Mini program of Osaka University. 
TS and YY acknowledge support from Thailand Science Research and Innovation and Suranaree University of Technology through the Royal Golden Jubilee Ph.D. Program (Grant No. PHD/0041/2555).
The work of SS is supported in part by the Rotary Yoneyama Memorial Foundation.
AH is supported in part by Grants-in Aid for Scientific Research on Innovative Areas, No. 18H0540.
\label{sec-6}

\end{document}